\documentclass[12pt]{iopart}

\usepackage{iopams}
\usepackage{color}
\usepackage[english]{babel}

\usepackage[pdftex]{graphicx}% Include figure files

\definecolor{myblue}{rgb}{0.2,0.2,0.8}
\definecolor{myblack}{rgb}{0,0,0}

\usepackage[colorlinks=true,citecolor=myblue,linkcolor=myblack]{hyperref}

 \newcommand{\ket}[1]{|#1\rangle}
 \newcommand{\bra}[1]{\langle #1|}
 
\def\oned{\mathrm{1d}}

\def\dd{\mathord{\rm d}}
\def\ee{\mathord{\rm e}}
\def\ii{\mathord{\rm i}}
\newcommand{\rs}{\mathrm{s}}
\newcommand{\rd}{\mathrm{d}}
\newcommand{\rg}{\mathrm{g}}

\newcommand{\ry}{\mathrm{y}}
\newcommand{\rt}{\mathrm{t}}
\renewcommand{\aa}{\mathrm{a}}

\newcommand{\src}{\mathrm{(s)}}
\newcommand{\trg}{\mathrm{(t)}}
\newcommand{\dtc}{\mathrm{(d)}}

\newcommand{\hps}{b_{s}}
\newcommand{\hpe}{b_{e}}
\newcommand{\hpsi}[1]{b_{s,#1}}
\newcommand{\hpei}[1]{b_{e,#1}}

\begin{document}

\title[Heralded multiphoton states with coherent spin interactions in waveguide QED]{Heralded multiphoton states with coherent spin interactions in waveguide QED}

\author{V. Paulisch$^1$, A. Gonz\'alez-Tudela$^1$, H. J. Kimble$^{2,3}$ and J.~I.~Cirac$^1$}

\address{$^1$ Max-Planck-Institute of Quantum Optics, Hans-Kopfermann-Strasse 1, 85748 Garching, Germany}
\address{$^2$ Norman Bridge Laboratory of Physics 12-33, California Institute of Technology, Pasadena, CA 91125, USA}
\address{$^3$ Institute for Quantum Information and Matter, California Institute of Technology, Pasadena, CA 91125, USA}

\ead{vanessa.paulisch@mpq.mpg.de}
\ead{alejandro.gonzalez-tudela@mpq.mpg.de}

\begin{abstract}

Waveguide QED offers the possibility of generating strong coherent atomic interactions either through appropriate atomic configurations in the dissipative regime or in the bandgap regime. In this work, we show how to harness these interactions in order to herald the generation of highly entangled atomic states, which afterwards can be mapped to generate single mode multi-photonic states with high fidelities. We introduce two protocols for the preparation of the atomic states, we discuss their performance and compare them to previous proposals. In particular, we show that one of them reaches high probability of success for systems with many atoms but low Purcell factors.

\end{abstract}

% \pacs{}

\vspace{2pc}
% \noindent{\it Keywords}:

% \submitto{\njp}
 
\maketitle

\section{Introduction}

Non-classical states of few photons can be generated in a variety of physical systems.
Triggered single photon-sources \cite{migdall_book13} can be found in solid state systems \cite{kurtsiefer00,zwiller03,pechal14}, in neutral atoms or ions coupled to optical cavities  \cite{law97,keller04,beugnon06,thompson06,barros09,muecke13} and in collective atomic ensembles \cite{chou04,chen06,simon07,eisaman04,ourjoumtsev06,bimbard10,lang13,cooper13}. By combining these single-photon sources with linear optics tools and post-selection, it is possible to achieve higher photon number states but with an exponentially small probability, which precludes the generation of larger photon numbers  (see \cite{dellanno06} and references therein) and, typically, destroying the state after heralding. 

The enhanced light-matter interactions provided by waveguide QED~\cite{laucht12,lodahl15,yu14a,thompson13a,tiecke14a,goban15a,vetsch10a,goban12a,petersen14a,beguin14a, sorensen16, corzo16} presents an excellent arena for the generation of multi-photon states. One possibility is to use atom-like metastable states of atom-like systems as quantum memories that can afterwards be triggered to generate photonic states with controllable temporal shape \cite{porras08a,gonzaleztudela15a} and with a very favorable scaling of the infidelity $I_\mathrm{phot} \propto m^2/(N P_\oned)$, with $N$ being the number of quantum emitters and $P_\oned$ the (single atom) Purcell Factor of the system which characterizes how much emission goes into the waveguide with respect to free space emission. Hence, the generation of arbitrary photonic states reduces to the preparation of arbitrary symmetric excitations in an ensemble of quantum emitters, which is the main focus of this work. 

In recent proposals \cite{gonzaleztudela15a,gonzaleztudela16a}, we designed both deterministic and probabilistic methods to generate collective atomic states using equally spaced atoms within the purely dissipative waveguide regime. It was shown that this simple atomic configuration allows one to prepare collective atomic states of $m$ excitations with either the fidelity \cite{gonzaleztudela15a} or the heralding probability \cite{gonzaleztudela16a} deviating from unity only by a factor scaling with $P_\oned^{-1/2}$. The key resource of these protocols is the long-range dissipative coupling induced by the waveguide which enforces effective unitary dynamics through the Quantum Zeno effect.

\begin{figure}[tb]
	\centering
	\includegraphics[width=0.99\textwidth]{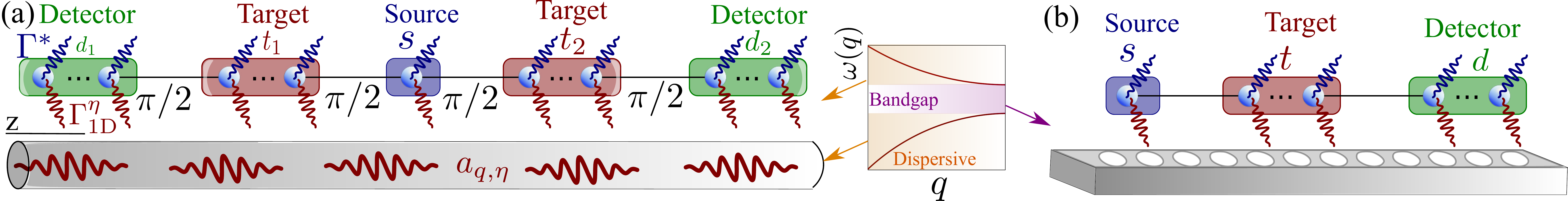}
	\caption{Ensembles of emitters coupled to one-dimensional reservoirs. The common ingredient is to have an individually adressable \emph{source} atom, used to transfer single excitations to the \emph{target} ensemble, which we herald with a change of state in the \emph{detector} atoms. (a) \emph{Double mirrors} configuration to achieve strong coherent interactions in the dissipative regime. (b) Engineered dielectric configuration which gives rise to strong coherent interactions in the bandgap regime.}
	\label{fig1}
\end{figure}

In this work, we present two protocols that harness long-range \emph{coherent} interactions induced by the guided modes to generate collective atomic excitations within an ensemble of atoms.
\begin{itemize}
\item The first protocol (``double mirrors'') is designed for emitters, whose resonance frequency corresponds to some guided mode in the band, see Figure \ref{fig1}(a), and which is, e.g., well suited for optical fiber setups. In this case, the atomic configuration determines the coherent and the dissipative interaction between the emitters. The atomic configuration in our protocol was inspired by \cite{chang12a}, in which a pair of atomic mirrors are placed next to a single emitter and where the analogy to cavity QED with $g \propto \sqrt{N}$ was shown. By placing another set of mirrors around this atomic cavity, we obtain the tools necessary for the heralded generation of collective atomic excitations in the first pair of mirrors.
\item The second protocol (``dipole-dipole'') is designed for emitters, whose resonance frequency is in the bandgap, see Figure \ref{fig1}(b), and which is, e.g., well suited for engineered dielectrics. In this regime, dipole-dipole interactions mediated by an atom-photon bound state formed in the bandgap emerge \cite{john90a,bykov75a,kurizki90a,douglas15a,gonzaleztudela15b}. Here, the advantage is that dissipation (through the waveguide modes) is strongly suppressed.
\end{itemize}
We analyze in detail the performance of both protocols, and compare it to previous ones.

The rest of the manuscript is divided as follows: in Section \ref{sec:System}, we introduce the two configurations and explain how the coherent coupling emerges in each case together with the common ideas of the protocol. In Section \ref{sec:dissipative}, we discuss the situation for the \emph{double mirrors} setup, explaining one main protocol, together with different variations of it. In Section \ref{sec:bandgap}, we discuss how to adapt the protocol in the dissipative regime for the situation where dipole-dipole interactions are mediated by atom-photon bound states. Finally, in Section \ref{sec:comparison} we summarize the figures of merit and scaling of the different protocols and make a comparison with previous proposals \cite{gonzaleztudela15a,gonzaleztudela16a}.

% 

%===============================================================================
\section{System and general protocol}\label{sec:System}

\begin{figure}[ht]
	\centering
	\includegraphics[width=0.99\textwidth]{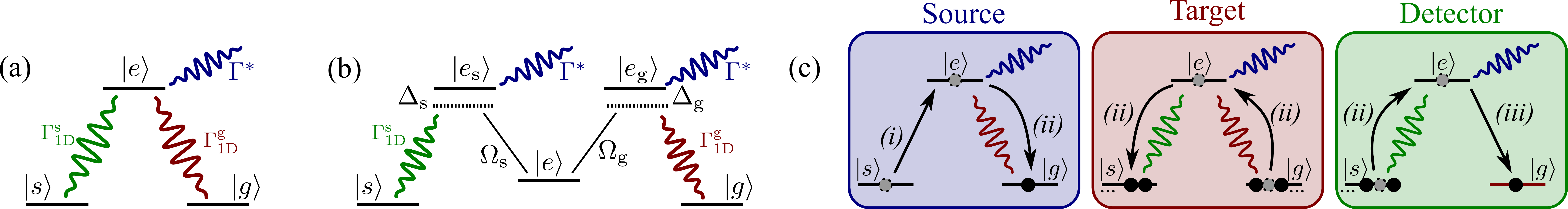}
	\caption{(a) Simplest internal level structure of emitters, in which two transitions in a three-level system are coupled to the waveguide.
		(b) Internal level structure with control over the effective decay rates to the waveguide, $\Gamma_\oned^\eta \frac{|\Omega_\eta|^2}{4\Delta_\eta^2}$, and to free space, $\sum_{\eta} \Gamma^* \frac{|\Omega_\eta|^2}{4\Delta_\eta^2}$, where $\Omega_i$ is the Rabi coupling strength between $\ket{e}$ and $\ket{e_i}$, see \cite{SupMat}. (c) Protocol for the repeated heralded addition of single excitations to the target ensemble. The heralding measurement is performed on the $\ket{g}$-state of the detector ensemble.}
	\label{fig2}
\end{figure}

As shown in Figure~\ref{fig1}, the common ingredient for both the dissipative and bandgap regime is to have three individually addressable ensembles, namely, a \emph{source} atom, used to transfer single excitations to the \emph{target} ensemble, which we herald with a change of state in the \emph{detector} ensemble. Moreover, the emitters must have two dipole transitions $\ket{g}\leftrightarrow \ket{e}$ and $\ket{s}\leftrightarrow \ket{e}$ coupled to two waveguide modes\footnote{For the dissipative regime, the waveguide mode is propagating, whereas in the bandgap regime an atom-photon bound state is formed. In both cases we assume the propagation length of the modes to be much larger than the system size.} as shown in Figure \ref{fig2}(a-b) in which either coupling can be controlled, e.g., by using a $M$-type level scheme (see Figure \ref{fig2}(b) or by Stark-shifting the respective levels out of resonance. The two guided modes are required so that the source emitter and the detector ensemble can couple to different modes. This ensures that no direct excitation transfer 
between them can take 
place. An excitation in the source emitter can only excite the detector atoms when a collective excitation is created in the target ensemble. Furthermore, we require in both cases that the source/target/detector ensembles are individually addressable.

Apart from these common ingredients, the two protocols in the dissipative and bandgap regime require additional conditions. For example, in the dissipative regime we demand that:
\begin{enumerate}
 \item The two guided modes mediating the interaction have equal wavelength $\lambda_0$, defined by the characteristic atomic frequency: $q(\omega_{es}) = q(\omega_{eg}) \equiv q_0=2\pi/\lambda_0$. This can be achieved, e.g., by the use of two different polarization modes.
 \item The emitters inside the target/detector ensemble are placed at distances commensurate with $\lambda_0$, whereas the source/target and target/detector are placed at a distance $\lambda_0/4 + n \lambda_0$ with $n \in \mathbb{N}_0$.\footnote{In fact, the protocol also works for distances within an ensemble of $\lambda_0/2$ (or multiples thereof) and/or distances between source/target or target/detector of odd multiples of $\lambda_0/4$ if the collective external drivings are adjusted accordingly.}
 \item We can neglect finite propagation lengths and non Markovian effects and thus use a Markovian description to analyze the performance. The impact of these effects has been discussed in References \cite{guimond16,shi15,paulisch16}. This requires that the maximal distance between atoms is small compared to the propagation length, $v_\mathrm{g} T$, during the time of operations, where $v_\mathrm{g}$ is the group velocity in the waveguide.
\end{enumerate}

For the configuration within the bandgap regime, the only additional requirement is the existence of a bandgap for the two guided modes, such that their interactions can be mediated by virtual guided modes.

%------------------------------------------------------------------------------- 
\subsection{Theoretical description in the dissipative regime}

In the dissipative regime we use what we denote as \emph{double mirrors} configuration as sketched in Figure~\ref{fig1}(a), in which the \emph{source} atom is embedded by two atomic cavity mirrors with $N$ atoms each which play the role of target atoms. Moreover, we embed the source/target system within two other atomic mirrors with $N$ atoms each, that altogether form the detector ensemble. When tracing out the reservoir degrees of freedom, we obtain an effective master equation which describes the atomic dynamics. In this configuration, the waveguide induces strong and long-range coherent spin interactions between the different ensembles described by the Hamiltonian \cite{SupMat}
\begin{equation}
\label{eq:H_dd}
H_\mathrm{wg} 
= \frac{\Gamma_\oned^\rg}{2}  \sigma_{ge}^\src S_{eg,+}^\trg 
+ \frac{ \Gamma_\oned^\rs }{2} S_{es,-}^\dtc S_{se,-}^\trg 
+ \mathrm{h.c.},
\end{equation}
where $S_{\alpha\beta,\pm}^{(\ry)} = S_{\alpha\beta}^{(\ry_1)} \pm  S_{\alpha\beta}^{(\ry_2)}$ are the collective operators within each ensemble $S_{\alpha\beta}^{(\mathrm{y})} = \sum_{j \in \mathrm{y}} \sigma_{\alpha \beta}^j$, for the target ensembles $\ry=\rt$ and for the detector ensembles $\ry=\rd$, and where $\sigma_{\alpha \beta}^j = \ket{\alpha}_j \bra{\beta}$. The renormalized spontaneous emission rate of the atoms, i.e., $\Gamma_{\oned}^\eta=\sum_q |g_q^\eta|^2 \delta(\omega_{e\eta}-\omega_{q,\eta})$, depends on both the energy dispersion ($\omega_{q,\nu}$) and guided mode profile (contained in $|g_{q}^{\eta}|$) \cite{goban13a}. Together with the coherent spin interactions also collective decay terms emerge, which are given by
\begin{equation}
\label{eq:Ld}
\mathcal{L}_{\mathrm{d}}[\rho]
= \frac{\Gamma_\oned^\rg}{2}D_{ \sigma_{ge}^\src}[\rho]
+\frac{\Gamma_\oned^\rg}{2}D_{ S_{ge,-}^\trg}[\rho]
 +\frac{\Gamma_\oned^\rs}{2}D_{ S_{se,-}^\trg}[\rho]
+\frac{\Gamma_\oned^\rs}{2}D_{ S_{se,+}^\dtc}[\rho],
\end{equation}
where $D_O[\rho] = 2 O \rho O^\dagger - O^\dagger O \rho - \rho O^\dagger O $. Obviously, apart from the decay into the desired waveguide modes, the excited states $\ket{e}_n$ may also emit photons into free space, or even to other non-guided waveguide modes. We include all these processes into a rate, $\Gamma^*$,  typically of the order of the natural linewidth $\Gamma_{\aa}$, which can be described through an additional Lindblad term in the Liouvillian as
\begin{equation}
\label{mequation1}
{\cal L}_{*}(\rho) = 
\sum_{i,\eta} \frac{\Gamma^*}{2} \left( \sigma^i_{\eta e} \rho \sigma^i_{e \eta} - \rho \sigma^i_{e \eta} \sigma^i_{\eta e} \right)
+ \rm {h.c.}\, ,
\end{equation}
which gives rise to the so-called Purcell Factor $P_{\oned}=\Gamma_{\oned}^{\rg}/\Gamma^*$, and where the index $i$ denotes the different atoms and $\eta$ the different hyperfine level where quantum jumps occur. Another important parameter is the ratio of decay rates $\Gamma_\oned^\rs / \Gamma_\oned^\rg$ which may be tunable or not, depending on the particular setup.

The source/target configuration is inspired by Reference \cite{chang12a}, where it was shown that this model can be mapped to a cavity QED configuration, where the source atom plays the role of the two-level system with effective decay $\gamma=\Gamma_\oned^\rg+\Gamma^*$ which couples coherently to an effective cavity defined by $\frac{1}{\sqrt{2N}} S^\trg_{eg,+}$, with rate $g=\sqrt{2N}\Gamma_\oned^\rg$, as shown in Equation~\ref{eq:H_dd} which is subradiant, i.e.,  the cavity loss is given by $\kappa=\Gamma^*$. It is known \cite{law96a} that within the strong non-linear coupling regime, e.g.,  when $g\gg \kappa,\gamma$, this model can generate deterministically any arbitrary superposition of the cavity-like mode up to $m$ photons with an error $\varepsilon\approx \frac{m (\kappa+\gamma)}{g}\propto \frac{m}{\sqrt{N}}$. In standard cavity QED, one could improve the scaling to $\varepsilon\sim 1/\sqrt{C}$, using off-resonant Raman transitions \cite{sorensen03a}, where $C=g^2/(\kappa \gamma)$ is the so-called cooperativity. In the effective model within the waveguide setup, 
however, 
both $\Gamma_\oned^\rg$ and $\Gamma^*$ (and consequently the effective cavity QED parameters $g,\kappa,\gamma$) renormalize in the same way when using off-resonant transitions such that the optimization is not possible.

\subsection{Theoretical description in the bandgap regime}

If we assume that both the $e-g$ and the $e-s$ transitions are within the bandgap regime, it can be shown \cite{douglas15a,gonzaleztudela15b} that the excited atomic states are dressed by a photon cloud of size $\xi d$ which allows to exchange interaction between the emitters, which are assumed to have an equidistant spacing $d = 2\pi / a$, where $a$ is the period of the photonic crystal. This distance is well-suited for the generation of photonic states in a later step and also avoids the sign alternation due to the phase acquired by the Bloch mode at the cut-off frequency \cite{albrecht16}. The photon cloud can be seen as an off-resonant atom-induced cavity of length $\xi d$, that allows to exchange interactions between the emitters described by the Hamiltonian
\begin{equation}
\label{eq:H_ddband}
H_\mathrm{bg} 
= \frac{\Gamma_\oned^\rg}{2\xi}  \sum_{n} e^{-\frac{|z^\src-z_n^\trg|}{\xi d}}\sigma_{ge}^\src \sigma_{n,eg}^\trg 
+ \frac{ \Gamma_\oned^\rs }{2 \xi} \sum_{n,m}  e^{-\frac{|z_n^\trg-z_m^\dtc|}{\xi d}} \sigma_{n,es}^{\dtc} \sigma_{m,se}^\trg 
+ \mathrm{h.c.}\,
\end{equation}
where $\Gamma_\oned^{\rg,\rs}$ are the decay rates at the bandgap frequency cutoff, $e^{-|z_n-z_m|/\xi d}$ is the overlap between the effective cavity of the $n$-th atom with the $m$-th atom, and the $1/\xi$ dependence is the decrease of the coupling strength to the cavity due to the increase of the mode length. Notice that apart from $H_\mathrm{bg}$, the photon cloud also induces dipole-dipole couplings within the target (and detector ensemble). We did not write them here explicitly, because their effect on our protocols can be compensated with appropriate laser detunings as will be explained in Section \ref{sec:bandgap}. In the limit of $L/d \gg \xi \gg N$, where $L$ is the length of the photonic crystal, the Hamiltonian of Equation~\ref{eq:H_ddband} converges to the one of Equation~\ref{eq:H_dd} with a renormalized $\Gamma_\oned^{\rg,\rs}$ by the factor $1/\xi$ but with the advantage of eliminating the collective quantum jumps of Equation~\ref{eq:Ld}. 

%-------------------------------------------------------------------------------
\subsection{Protocol}

The basic principle of our protocol for both the dissipative and the bandgap regimes is depicted in Figure \ref{fig2}(c). In order to add one (symmetric) excitation to the \textit{target} ensemble already containing $m-1$ excitations, (i) the \textit{source} atom is excited, (ii) by dipole-dipole coupling the excitation is collectively transfered to the target ensemble through the $\Gamma_\oned^\rg$-mode, and further to the \textit{detector} ensemble through the $\Gamma_\oned^\rs$-mode, and finally (iii) a fast $\pi$-pulse on the detector ensemble's $\ket{g}\leftrightarrow\ket{e}$-transition terminates the dynamics. If a heralding measurement on the $\ket{g}$ state on the emitters of the detector ensemble is successful, a symmetric excitation in the target ensemble must have been generated with a heralding probability that we denote as $p_{m-1\rightarrow m}$ and an error or infidelity $I_{m-1\rightarrow m}$. If a collective excitation has been added, the source emitter and the detector ensemble are 
reinitialized and the 
process is repeated. Hence, to reach any state with $m$ (symmetric) excitations in the target ensemble, one will have to repeat the above procedure successfully $m$ times. 

If at some point a heralding measurement fails, either a quantum jump in one of the ensembles has occurred or the excitation has not been transfered to the detector ensemble yet. Some of these processes do not spoil the coherence of the target ensemble state and would be correctable. However, because all these processes are indistinguishable, the whole protocol has to be repeated from the very beginning to avoid a low fidelity of the final state, that is, a low overlap with the target state. The final protocol to accumulate $m$ excitations will be characterized by the average number of operations $R_m=\left(\prod_{k=1}^m p_{k-1\rightarrow k}\right)^{-1}$, which is in general exponential in $m$, and has a total infidelity
\begin{equation}
	I_m=1-\sqrt{\langle \psi_\mathrm{t}| \rho | \psi_\mathrm{t} \rangle},
	\label{eq:Im}
\end{equation}
where $\ket{\psi_\mathrm{t} }$ is the target state and $\rho$ is the actual final state. 

In Section \ref{sec:dissipative}, we first discuss in detail the protocol in the dissipative regime and will find an ultimate limit that is imposed by the collective quantum jumps of Equation \ref{eq:Ld}. Then, in Section \ref{sec:bandgap} we discuss how to adapt the protocol to the case of finite range of the dipole-dipole couplings $\xi d$ and the limitations imposed by it.

%===============================================================================
\section{Detailed protocol in the dissipative regime}\label{sec:dissipative}

The practicality of the outlined protocol is gauged by the heralding probability and the fidelity of the final state with respect to the target state, $\ket{\psi_\mathrm{t}} = \sum c_m \ket{\psi_m}$ with $\ket{\psi_m} \propto S_{sg,-}^m \ket{\psi_0}$, where the initial state is $\ket{\psi_0} = \ket{g^\trg}^{\otimes 2N}$. Because the Hamiltonian $H_\mathrm{wg}$ of Equation (\ref{eq:H_dd}) leaves the excitation number $m$ invariant and because the state is heralded at the end of every cycle, we only need to treat the case in which one excitation is added to the state of the target ensemble $\ket{\psi_{m-1}}$. If the heralding measurement is successful, the state is then $\ket{\psi_{m}}$. The full initial state of the system is denoted by $\ket{\phi_0^m} = \ket{s^\src}\otimes\ket{\psi_{m-1}}\otimes\ket{s^\dtc}^{\otimes 2N}$. In the following analysis, we will skip most of the technical details and refer the interested reader to the Appendix \cite{SupMat}.

%-------------------------------------------------------------------------------
\subsection{Holstein-Primakoff-Approximation: calculation of probability}

For large ensemble sizes, which are necessary for the photon generation step, the low excitation regime, i.e., $m\ll N$, is approximately bosonic. In this case, the multilevel Holstein-Primakoff Approximation (see \cite{holstein40a, kurucz10, SupMat}) can be applied.\footnote{In this case, also the mapping of the source and target ensemble to cavity QED is perfect.} Then, the spin operators in the ensembles $\rt_1$ and  $\rt_2$ are approximated by bosonic operators $b_{\eta,j}$ up to $\mathcal{O}\left(\frac{1}{N}\right)$ as
\begin{equation}
	S^\trg_{eg,1(2)} \approx \sqrt{N} \hpei{1(2)}^\dagger,\ S^\trg_{se,1(2)} = \hpsi{1(2)}^\dagger \hpei{1(2)}.
\end{equation}
Then, the Hamiltonian $H_\mathrm{wg}$ of Equation \ref{eq:H_dd} couples the initial state to two other normalized states, that is,
\begin{eqnarray}
	\sigma_{es}^\src \ket{\phi_0^m} 
		\stackrel{\sqrt{2N} \Gamma_\oned^\rg}{\longleftrightarrow }&
	\sigma_{gs}^\src \frac{1}{\sqrt{2}}(\hpei{1}^\dagger + \hpei{2}^\dagger) \ket{\phi_0^m} \nonumber \\
		&\stackrel{\sqrt{2N m } \Gamma_\oned^\rs}{\longleftrightarrow }
	\sigma_{gs}^\src \frac{1}{\sqrt{2m}}(\hpsi{1}^\dagger - \hpsi{2}^\dagger) \frac{1}{\sqrt{2N}} S_{es,-}^\dtc \ket{\phi_0^m}\,.
\end{eqnarray}

Other non-excited states that are reached by quantum jumps can be neglected because of the heralding step. The dynamics is then governed by the non-hermitian Hamiltonian $H_\mathrm{wg} - \frac{\ii}{2} \sum O_k^\dagger O_k$, where the sum runs over all Lindblad operators $O_k$ of the Liouvillian $\mathcal{L}_\mathrm{d}$ of Equation \ref{eq:Ld}.
The coupling strength between the states scales with $\sqrt{N}$ due to the enhanced coupling of the collective states and the scaling with $\sqrt{m}$ appears because the symmetric excitation of the target ensemble is superradiant with respect to the $\Gamma_\oned^\rs$-mode \cite{dicke54a}.

\begin{figure}[t]
	\centering
	\includegraphics[width=0.99\textwidth]{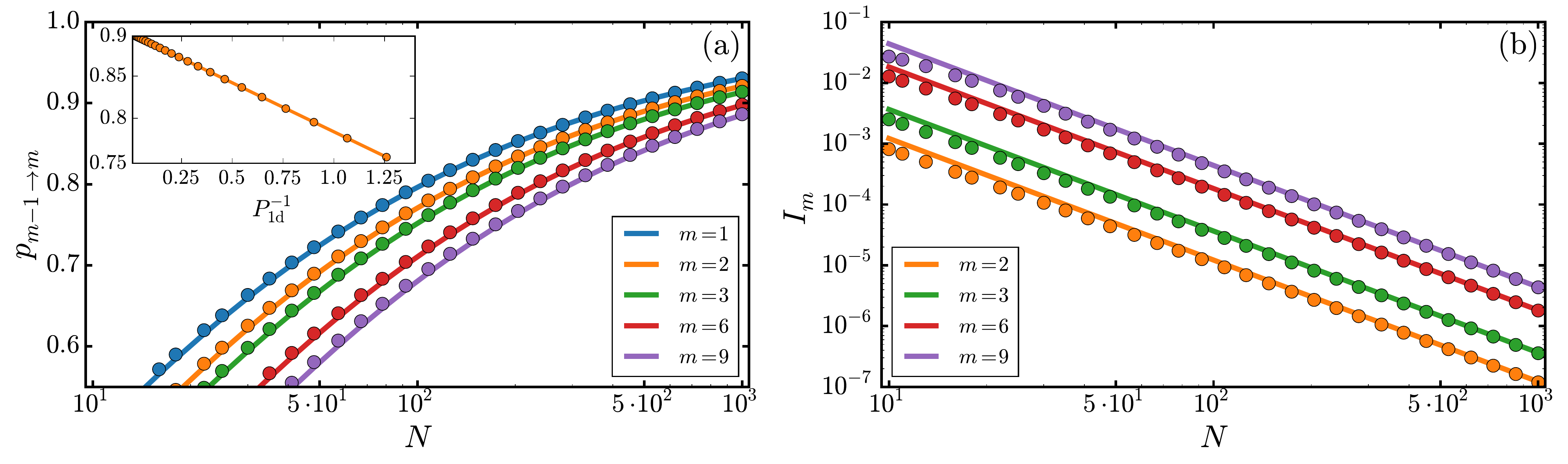}
	\caption{
		(a) For the success probability, Equation \ref{eq:p_m} is a good approximation (solid line) of the full solution (circles) with $P_\oned = 10$. The inset (a lin-log plot) shows the scaling with the Purcell Factor for $N=500$ and $m=2$, which satisfies $\ln(p_{m-1\rightarrow m}) \propto P_\oned^{-1}$.
		(b) The infidelity $I_{m}$ to accumulate $m$ excitations is shown as a function of $N$. Within the Holstein-Primakoff-Approximation unit overlap can be reached. The infidelity is independent of the Purcell Factor.}
	\label{fig:pSucc}
\end{figure}

A maximal population transfer of the excitation in the source to the detector ensemble is obtained for tunable coupling strengths, in particular, $\Gamma_\oned^\rg = \sqrt{m} \Gamma_\oned^\rs$, and for an evolution ending with a $\pi$-pulse on the detector ensemble after $T = \sqrt{2} \pi \left(\sqrt{2N} \Gamma_\oned^\rg \right)^{-1}$. Other choices for the parameters, in particular the ones allowing for $\Gamma_\oned^\rg = \Gamma_\oned^\rs$, may also lead to a sufficiently high success probability and fidelity. These variations are discussed at the end of this Section.
In the optimal case, the success probability of the heralding measurement (Figure \ref{fig:pSucc}a) is \cite{SupMat}
\begin{equation}
	p_{m-1\rightarrow m} \approx 
	\mathrm{exp} 
	\left[ 
	- \frac{\sqrt{2} \pi}{8 \sqrt{2N}} \left( 3 + 2 \sqrt{m} + 8 P_\oned^{-1} \right)
	\right].
	\label{eq:p_m}
\end{equation}
The scaling originates from the fact that the process is very fast, $T\propto N^{-1/2}$, and that the non-hermitian terms, which lead to the reduction of the success probability, scale with $\Gamma_\oned^\rg$ and $m \Gamma_\oned^\rs = \sqrt{m} \Gamma_\oned^\rg$. This enhanced decay rate can also be avoided by variations of this protocol. The prefactors in the exponential in front of $\Gamma_\oned^\rg T$ and $\sqrt{m} \Gamma_\oned^\rg T$ arise from the population of the specific states which are subject to the respective quantum jumps, that is $\frac{1}{T }\int_0^T \| S_{ee}^\src \ket{\psi(t)} \|^2 \dd t \approx \frac{3}{8}$ and $\frac{1}{T} \int_0^T \| S_{ee}^\trg \ket{\psi(t)} \|^2 \dd t \approx \frac{2}{8}$. The dependence on the Purcell Factor is exact because the evolution takes place in the subspace of a single excitation and every state is affected in the same way by spontaneous emission.

By repeatedly adding heralded single excitations, the state with $m$ collective excitations $\ket{\psi_m}$ can be reached. Clearly, the average number of repetitions is exponential in the number of excitations, i.e.,  $R_m=\prod_{k=1}^{m} p_{k-1\rightarrow k}^{-1} \propto e^{m\sqrt{m/N}} $, for $1\ll m\ll N$. Within the Holstein-Primakoff Approximation, no other states are coupled and therefore there would be no error. However, we will see that this is not true when one considers the corrections to the dynamics in the Holstein-Primakoff picture.

%-------------------------------------------------------------------------------
\subsection{Beyond Holstein-Primakoff approximation: calculation of fidelities.}

For the generation of a single excitation the Holstein-Primakoff Approximation is exact. However, for higher excitations, the non-hermitian part of the Hamiltonian leads to a coupling to additional states, that are linearly independent of the three states treated above. Also, the mapping of the source and target ensemble to cavity QED is no longer perfect and will suffer from a similar loss in fidelity as our protocol. The deviations from the approximation can be investigated numerically by using the exact Holstein-Primakoff Transformation (see Appendix). Instead of three orthonormal states as above, one then has to consider $4 m + 1$ orthonormal states, which are symmetric in each ensemble. For obtaining the results, the bosonic operators $\hpei{i}$ and $\hpsi{i}$ are cut-off at $2$ and $m+1$, respectively.

The multitude of additional states that the non-hermitian Hamiltonian couples to may lead to a non-unit overlap with the target state $\ket{\psi_m}$, which should go to unity in the limit of large ensemble size $N\gg1$.
Therefore, also the new initial state of the target deviates from $\ket{\psi_{m-1}}$ and has to be obtained from the final state of the preceding step.
The results from the full numerical analysis (Figure \ref{fig:pSucc}) agree very well with the results obtained by applying the Holstein-Primakoff-Approximation for $N \gg m$. Furthermore, the average accumulated infidelity as defined in Equation  (\ref{eq:Im}) is very close to unity and scales (for $m\ll N$) as
\begin{equation}
I_m \approx 0.061 \cdot \frac{m(m-1)}{N^2},
\label{fig:1-F}
\end{equation}
where the prefactor was obtained by a fit of the results from numerical integration of the master equation. The fidelity is independent of the Purcell Factor because every state is affected in the same way by spontaneous emission and the transitions to unwanted states only happens through collective operators.

\subsection{Variations of the protocols}

The protocol described in the previous Section used several requirements, e.g., tunable coupling to guided modes or fast $\pi$-pulses, to maximize the heralding probability while keeping the infidelity minimal. If some of these ingredients are not available there exist several alternatives to obtain still high heralding probabilities. For example,
\begin{itemize}
 \item \emph{Fixed coupling to waveguide modes.}  Typically, the Purcell factor $P_\oned \equiv \Gamma_\oned^\rg / \Gamma^*$ and the target ensemble size $N$ are fixed, leaving the ratio of decay rates $\Gamma_\oned^\rs / \Gamma_\oned^\rg$ and the final time $T$ at which the detector atoms are de-excited as the parameters open for optimization.\footnote{In principle, one can also consider the case in which the target ensemble and the detector ensemble are not of the same size $N$, but this does not lead to qualitatively different results.}
 
 When the decay rates have a fixed ratio, e.g., $\Gamma_\oned^\rg / \Gamma_\oned^\rs = 1$, a full population transfer to the detector ensemble is not possible. For large ensemble sizes, the heralding probability approaches
\begin{equation}
	\widetilde{p}_{m-1\rightarrow m}
	\stackrel{N\rightarrow \infty}{\longrightarrow} 
	\frac{4m}{(m+1)^2},
\end{equation}
see Appendix for details. Interestingly, the infidelity of Equation \ref{fig:1-F} is unchanged.

\item \emph{Replacing fast $\pi$-pulses.} The fast $\pi$-pulse on the source atom at the beginning and on the detector ensemble at the end of each step can be avoided by applying a continuous external field with the same Rabi coupling strength $\Omega$ to the respective transitions. These are the $\ket{s}-\ket{e}$-transition of the source atom and the $\ket{g}-\ket{e}$-transition of the detector ensemble. The success probability is then maximized for the same ratio of the decay rates, i.e., $\Gamma_\oned^\rg = \sqrt{m} \Gamma_\oned^\rs$, for $\Omega = \sqrt{\frac{2}{3}} \sqrt{2N} \Gamma_\oned^\rg$ and for $T= 3 \pi / \Omega$.
The scaling of the success probability with the parameters remains the same and only some prefactors in the exponent change slightly, so that
\begin{equation}
		\widetilde{p}_{m-1\rightarrow m} = \mathrm{exp}
		\left[-\frac{\sqrt{6} \pi}{\sqrt{2 N}} 
		\left( \frac{10 + 9 \sqrt{m}}{64} + \frac{29}{64} P_\oned^{-1} \right)\right].
	\end{equation}

\item \emph{Using only one guided mode.} Even if only a single guided mode is available, say the $\Gamma_\oned^\rg$-mode, the proposed protocol can still be applied if an additional metastable state $\ket{c}$ in the target ensemble is available to which spontaneous emission $\Gamma_\mathrm{c}^* \ll \Gamma^*$ is strongly suppressed \cite{enk97,borregaard15a}. The fidelity is then limited by the precision of a  $\pi$-pulse between the two metastable ground states $\ket{g}$ and $\ket{s}$ and the ratio $\Gamma_\mathrm{c}^* / \left(\sqrt{N} \Gamma_\oned^\rg \right)$ (see Appendix for details). 

\item \emph{Adding $m$ excitations at once.} Instead of generating single excitations in every step through a single source atom, one could in principle also use a source ensemble of size $m$ and transfer all excitations to the target ensemble to generate $m$ collective excitations at once. However, the source atoms are then superradiant and decay with an enhanced decay rate of at least $m \Gamma_\oned$. On the other hand, the dipole couplings are only enhanced by $\sqrt{m N }\Gamma_\oned$, which implies that the probability would still scale exponentially with $\sqrt{\frac{m}{N}}$. In addition, one requires a measurement device which can resolve the excitation number of the detector ensemble to guarantee the transfer of $m$ excitations to the target ensemble. Even if that is possible, e.g., the probability for generating two excitations at once is lower than the probability, $p_{0\rightarrow1} p_{1\rightarrow2}$, obtained through the original protocol (see Appendix).
\end{itemize}

In all of these variations, the final goal is to accumulate several excitations within the same hyperfine level $\ket{s}$. When the heralding fails, we reinitialize the process all over again, which yields an exponential number of operations $R_m$ with the number of excitations we want to create. Moreover, the existence of $m$ excitations already in the state $s$ causes the enhanced decay $m\Gamma_\oned^\rs$ of the target ensemble, which leads to a scaling of the success probability with $\sqrt{m}$ and to the necessity of a tunable ratio $\Gamma_\oned^\rg / \Gamma_\oned^\rs$ for maximizing the success probability. The former can be avoided if after each heralding of a single collective excitation,  it is stored in other hyperfine levels available $\ket{s_n}$ to combine them a posteriori  using Raman two-photon and microwave transitions plus atomic detection. It can then be shown \cite{gonzaleztudela16a,fiurasek05a,motes16} that by using only one additional hyperfine level, $s_1$, the number of operations is still 
exponential $R_m\propto e^m$, whereas, if we use $\log_2 m$ levels a subexponential scaling of $R_m$ can be achieved. In these cases, carefully constructed repumping schemes have to be used to avoid introducing additional errors during the repumping step \cite{gonzaleztudela16a}.

%===============================================================================
\section{Protocol in the bandgap regime}\label{sec:bandgap}

In the previous Section, we showed how the success probability of the protocol in the dissipative regime is limited by quantum jumps into the waveguide, which leads to the scaling with $1-p_{m-1\rightarrow m} \propto \sqrt{m/N}$. As we showed in Section \ref{sec:System}, a possibility to get rid of the quantum jumps while maintaining the dipole-dipole interactions is to use interactions mediated by the bandgap \cite{john90a,bykov75a,kurizki90a,douglas15a,gonzaleztudela15b}. This can be interpreted as the formation of an atom-photon bound state mediating the exchange of interactions between emitters.
By using this configuration we eliminate the quantum jumps into the waveguide at the price of reducing the dipole-dipole couplings due to their finite range $\xi d$. In principle, one can make $\xi d$ much larger than the characteristic length of the system, $N d$, so that all the emitters couple homogeneously as in the dissipative regime. However, this comes at a price of enlarging the length of the atom-induced cavity and therefore the subsequent reduction of the dipole-dipole coupling. In this Section, we first discuss the scaling of the success probability and infidelity in the ideal limit $\xi \gg N$. In realistic cases, $\xi d$ is limited by the length of the photonic crystal $L$, that is we require a finite $\xi d \ll L$. Thus, we also explore the limitations imposed by this trade-off to generate multipartite entangled states.

%-------------------------------------------------------------------------------
\subsection{Ideal Case}
The idea of the protocol is analogous to the one in the dissipative regime: transfer a single excitation from the source atom to the target ensemble through the $\Gamma_\oned^\rg$ mode, and then from the target to the detector through $\Gamma_\oned^\rs$. In the limit $\xi \gg N$ the Hamiltonian $H_\mathrm{bg}$ of equation \ref{eq:H_ddband} converges to
\begin{equation}
	H_\mathrm{bg} = 
	\frac{\Gamma_\oned^\rg}{2\xi}  
		\left(  \sigma_{eg}^\src S_{ge}^\trg + 
				S_{eg}^\trg \sigma_{ge}^\src \right) 
	+ \frac{\Gamma_\oned^\rs}{2\xi}
				\left(  S_{es}^\trg S_{se}^\dtc + 
				S_{es}^\dtc S_{se}^\trg  \right)
	 + H_\mathrm{LS},
\end{equation}
where the dipole-dipole couplings within the ensembles have been included in the Hamiltonian $H_\mathrm{LS}$. The dynamics of the system can be again best analyzed by using the Holstein-Primakoff-Transformation. The main differences to the dissipative regime are the following:
\begin{enumerate}
	\item The emitters within each ensemble suffer dipole-dipole interactions irrespective of their position, whereas in the dissipative regime these can be canceled by choosing the $2\pi$ (or $\pi$) distances. These dipole-dipole interactions for equidistantly spaced atoms lead to a collective Lamb-shift, which is, e.g., for the $S_{eg}^\trg$ mode equal to $\Delta_\mathrm{L} = \frac{N_m \Gamma_\oned^\rg}{2\xi}$, where $N_m = N-m+1$. Therefore, in order to make the coherent transfer of excitations resonant, this Lamb-shifts in each ensemble have to be compensated through either appropriate Stark-shifts or magnetic field gradients for the source atom, target and detector ensembles.
	
	\item  The effective coherent coupling $G$ between the source atom and the collective mode of the target ensemble is also modified by the effective cavity length $\xi d$ and is given by $G = \frac{\sqrt{N}\Gamma_\oned^\rg}{2\xi}$. Therefore, the optimal time for a full population transfer to the target ensemble is $T = \pi/G$, if the detector ensemble is neglected. If one takes into account the second step, the maximal population transfer to the detector ensemble occurs for tunable coupling strengths, in particular again for $\Gamma_\oned^\rg = \sqrt{m} \Gamma_\oned^\rs$, at a time $\sqrt{2} \pi / G$. 
\end{enumerate}

From the previous analysis, we infer that the success probability of adding another excitation in the optimal case is
\begin{equation}
	p_{m-1\rightarrow m} \approx \exp \left[ -\frac{\pi \xi}{\sqrt{N_m} P_\oned}\right],
\end{equation}
where $N_m = N-m+1$. This scaling orginates in the fact that the timescale of the transfer $T \propto \xi/(\sqrt{N_m}\Gamma_\oned^\rg) $, whereas the only process that makes the norm decay is the spontaneous emission probability with rate $\Gamma^*$.

In general, this scaling is not better than in the dissipative regime. But as we showed in the previous Section, the imperfect fidelity was arising from the collective quantum jump terms, which are vanishing in this case such that the infidelity with the final state satisfies $I_{m} = 0$.

%-------------------------------------------------------------------------------
\subsection{Realistic Case $\xi \sim L/d$}
So far we have neglected the effect of a finite effective cavity length $\xi d$ on the state itself. In order to avoid unnecessary overlap between Sections, we focus on analyzing the effect of finite $\xi$ in the first step, i.e., for the transfer of a single excitation from the source atom to the target ensemble in state $\ket{g}^{\otimes N}$. Even though, this does not grasp the full process, it gives insight into the scaling of the infidelity with the finite effective cavity length $\xi$. The simplified Hamiltonian in the bandgap regime (with explicit dipole-dipole shifts of the ensemble) for the first step is then
\begin{eqnarray}
	H
	= \frac{\Gamma_\oned^\rg}{2\xi} & \left[ \sum_{n}  
		\left( e^{-\frac{|z^\src-z_n^\trg|}{\xi d}}\sigma_{ge}^\src \sigma_{n,eg}^\trg + \mathrm{h.c.} \right) \right. \nonumber \\
	&\left.
	+ \sigma_{ee}^\src
	+ \sum_{n,m}  e^{-\frac{|z_n^\trg-z_m^\dtc|}{\xi d}} 
		\sigma_{n,eg}^{\trg} \sigma_{m,ge}^\trg \right]
\end{eqnarray}

\begin{figure}[t]
	\centering
	\includegraphics[width=0.95\textwidth]{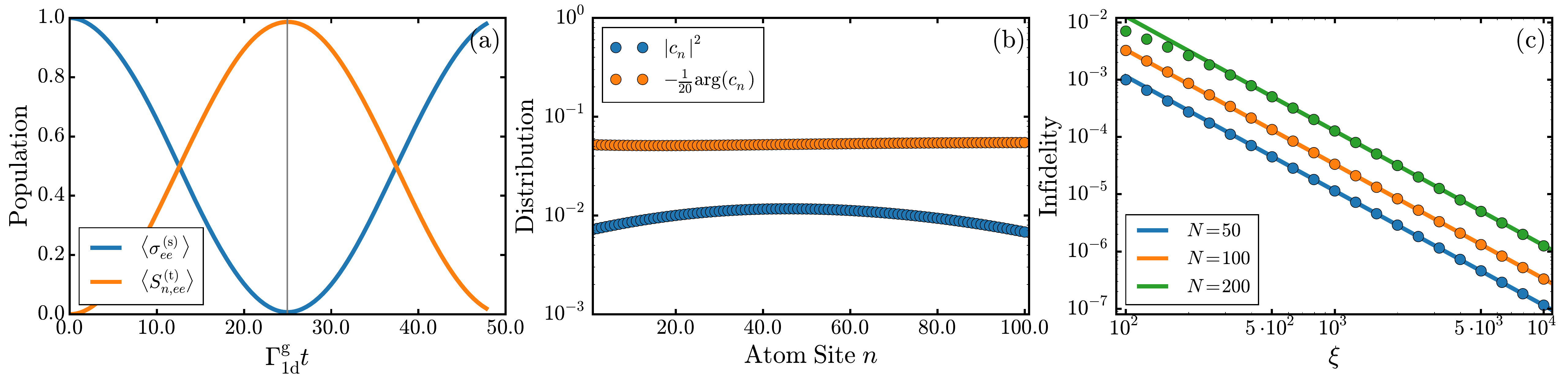}
	\caption{
		(a) Population dynamics of the coherent transfer of a collective excitation from source (blue) to a collective mode of the target ensemble (orange) for a situation with $N=100=\xi$ and $m=1$ and $\Gamma^*=0$.
		(b) Intensity and phase distribution of the target ensemble as defined in the text at the optimal time of the transfer for the same values as in (a).
		(c) Scaling of the infidelity of the intermediate state with respect to the fully symmetric state $S_{eg}^\trg \ket{g}^{\otimes N}$ after projection onto the subspace of states in which the excitation is in the target ensemble. The solid lines are a fit to the numerical values showing a scaling with $\xi^{-2}$.
		}
	\label{fig:Reactive}
\end{figure}

For finite $\xi$ one needs to take into account the changes in the collective Lamb-shifts and the coherent couplings. When taking this into account, the dynamics still lead to a (almost full) depletion of the population in the excited state of the source atom, see Figure \ref{fig:Reactive}(a) for $N=100$ and $\xi=100$. At the point of maximal population transfer we plotted the phase $\arg(c_n)$ and the intensity distribution $|c_n|^2$ for the coefficients $c_n$ of $\sigma_{n,eg}^\trg \ket{g^\trg}^{\otimes N}$ in the target ensemble in Figure \ref{fig:Reactive}(b). We see that in spite of the limited range $\xi\sim N$, the collective mode is approximately homogeneous. For smaller $\xi$, the phase and intensity distributions become inhomogeneous and therefore cannot be used to transfer the excitations coherently. In Figure \ref{fig:Reactive}(c) we show the scaling of the infidelity of the intermediate state with respect to the completely symmetric state and see that the infidelity scales favorably with the cavity length, i.e., as $I \propto \xi^{-2}$.

%===============================================================================
\section{Comparison between different protocols.}\label{sec:comparison}

\begin{table}[b]
		\begin{tabular}{l||cc|l}
			Protocol & Error $1-F$ & Success Probability $p_{m}$ & Requirements   \\
			\hline
			Deterministic \cite{gonzaleztudela15a} & $m / \sqrt{P_\oned}$ & $1$ & $P_\oned \gg 1$
			\\
			Probabilistic I \cite{gonzaleztudela16a}& $m (1-\eta) x^2$ & $(\eta x^2)^m$ & $x = \Omega T \sqrt{N} \ll 1$
			\\
			Probabilistic  II \cite{gonzaleztudela16a} & $0$ & $\ee^{-m / \sqrt{P_\oned}}$ & $P_\oned \gg 1 $
			\\
			``Double Mirrors'' & $m^2 / N^2 $ & $\ee^{-m \sqrt{\frac{m}{N}}(1+P_\oned^{-1})}$ & $N\gg 1$
			\\
			``Dipole-Dipole'' & $\mathcal{O}(\xi^{-2})$ & $\ee^{-\xi / (\sqrt{N_m} P_\oned)}$ & $\xi \gg N $
	\end{tabular}
	\caption{Summary of different protocols with the scaling for the accumulated error and the success probability for large $m$. See text for details of the protocols. For simplicity we leave out various constants to make clear the scaling with the relevant parameters.}\label{tab:comparison}
\end{table}

The protocols presented in this manuscript together with the ones presented in References~\cite{gonzaleztudela15a,gonzaleztudela16a} constitute a set of methods for quantum state preparation using different resources present in waveguide setups. To give a full understanding, we summarize the conditions and figures of merit for each protocol, identifying which ones are more suitable depending on the available resources (see Table \ref{tab:comparison}):
\begin{itemize}
 \item In Reference \cite{gonzaleztudela15a}, we use atomic $\Lambda$-systems with equally spaced atoms to build up arbitrary superpositions of atomic/photonic states. The protocol requires $P_\oned \gg 1$, is \emph{deterministic} and its infidelity to generate up to $m$ excitations scales as $I_m \approx m/\sqrt{P_\oned} + \mathcal{O}(P_\oned^{-1})$. This protocols is well suited for engineered dielectrics or in general any system with $P_\oned \gg 1$ \cite{laucht12,lodahl15,yu14a,thompson13a,tiecke14a,goban15a}. They can also be extended to low mode cavity QED systems if the same conditions hold, i.e., if one works in the bad-cavity limit and has an ancilla atom which can be addressed individually.
 
 \item The first protocol discussed in \cite{gonzaleztudela16a} also uses $\Lambda$-systems, requires $N P_\oned \gg 1$ and the use of an external single photon detector. The protocol heralds (by a photodetector with efficiency $\eta$) the transfer of single collective excitations with probability $p$, which can be controlled at will, but with a trade-off with the infidelities, which scales as $I_{m-1\rightarrow m}\propto p$. To accumulate excitations, the average number of operations is exponential $R_m\propto p^{-m}$, which hinders its extension for very large excitation numbers. However, it is especially suited for generating low photon numbers in systems with either large $P_\oned$ or systems with $P_\oned<1$ and large atom number $N$ like current experimental setups for optical fibers \cite{vetsch10a,goban12a,petersen14a,beguin14a,sorensen16, corzo16}.
 
 \item The other protocols discussed in \cite{gonzaleztudela16a}, also exploit the long-range dissipative coupling for equally spaced atoms, and require $P_\oned \gg 1$. The advantage is that the probability of heralding a single collective excitation $p\propto e^{-\sqrt{P_\oned}}$ can be made close to $1$ for systems with $P_\oned\gg 1$. Moreover, the infidelity of accumulating $m$ excitations is strictly $I_m=0$. This is certainly the best suited method in terms of fidelities but to obtain high probabilities we require systems with $P_\oned\gg 1$.
 
 \item The protocol within the dissipative regime discussed along this manuscript allows to overcome the limitations of probabilities for systems with low $P_\oned$, by putting the difficulty in a more elaborate configuration of atomic positions (see Figure~\ref{fig1}a). The heralding probability of a single collective excitation is only dependent on $N$, i.e., $p\propto e^{-\sqrt{m/N}}$. The average infidelity to accumulate $m$ excitations, though not being 0, is still very small, $I_{m}\propto m^2/N^2$. This is probably the best method for optical fiber setups \cite{vetsch10a,goban12a,petersen14a,beguin14a,sorensen16, corzo16}.
 
 \item The protocol within the bandgap regime is only suited for engineered dielectrics where the existence of bandgaps is possible. Though it has the advantage of eliminating quantum jumps into the waveguide, the finite range of the interactions $\xi$, leads to a worse heralding probability scaling with $p\propto e^{-\xi/(\sqrt{N_m}P_\oned)}$ with the advantage of a possibly better infidelity $I_m = \mathcal{O}\left(\xi^{-2}\right)$ in the ideal regime where the length of the photon-bound state mediating the interaction is larger than the ensemble size.
  
\end{itemize}

%===============================================================================
\section{Conclusions}\label{sec:Summary}

In conclusion, we have proposed several methods for the heralded preparation of symmetric states in ensembles of emitters using coherent atom-atom interactions induced by their coupling to two guided modes in waveguide QED setups in both the dissipative and the bandgap regime. In the dissipative regime, we showed how the collective quantum jumps into the waveguide limit the single excitation heralding probability $p_{m-1\rightarrow m} \geq e^{-\pi \sqrt{m/N}}$ which can still be close to $1$ for systems even with $P_\oned <1$, which is very relevant for the experiments with optical fiber setups \cite{vetsch10a,goban12a,petersen14a,beguin14a}. We also consider the situation of enginereed dielectrics within the bandgap regime in which the finite range of atom-induced cavities gives a more limited scaling of the probabilities. In all cases by using atomic detection and post-selection we rule out most of the errors, giving rise to very low global infidelities in both the dissipative ($I_m \propto m^2/N^2$) and 
bandgap regimes ($I_m =0$, for $\xi\gg N$) for the preparation of atomic states.

These prepared states can then be mapped to a photonic state of the waveguide with controllable temporal shape \cite{porras08a,gonzaleztudela15a}. This mapping scales favorably with the system parameters, in particular the emitter number $N$ and the Purcell factor $P_\oned$, that is the infidelity (or error) of this process scales as $I_\mathrm{ph} \propto m^2/(N P_\oned)$. Therefore, this protocol can be used for the efficient preparation of triggered multiphoton states.

%-------------------------------------------------------------------------------
\ack

VP acknowledges the Cluster of Excellence NIM.
AGT also acknowledges support from Intra-European Marie-Curie Fellowship NanoQuIS (625955).
HJK acknowledges funding by the Air Force Office of Scientific Research, Quantum Memories in Photon-Atomic-Solid State Systems (QuMPASS) Multidisciplinary University Research Initiative (MURI) by NSF PHY1205729, by the Institute of Quantum Information and Matter, a National Science Foundation (NSF) Physics Frontier Center with support of the Moore Foundation, and support as a Max Planck Institute for Quantum Optics Distinguished Scholar.

\appendix
\setcounter{section}{1}
\setcounter{equation}{0}

\section*{Appendix A: Atomic Dynamics in Dissipative Regime \label{sec:apA}}

\subsection{Derivation of the Master Equation}

Under the assumptions described in the main manuscript, the interaction Hamiltonian between the emitters and the one-dimensional reservoir takes the form ($\hbar = 1$)
\begin{equation}
H_\mathrm{I}  = \sum_{n,q,\eta} g_q^\eta \left( \sigma_{e\eta}^n a_{q,\eta} e^{i q z} + \rm {H.c.} \right),
\label{Hint}
\end{equation}
where the sum runs over all emitters and where $g_q^\eta$ is the single photon coupling constant for the dipole transition $\ket{e} \leftrightarrow \ket{\eta}$, which is independent of the atomic position. The atomic operator is denoted by $\sigma_{\alpha\beta}^{n} = \ket{\alpha}_n\bra{\beta}$ and the photonic annihilation operator for the respective mode is $a_{q,\eta}$. The one-dimensional bosonic reservoir is described by $H_{\rm b} =\sum_{q,\eta} \omega_{q,\eta} a^\dagger_{q,\eta} a_{q,\eta}$, where $\omega_{q,\eta}$ is the field dispersion relation. The atomic free energy of the dipole transition is given by $H_\aa = \omega_e \sum_j \sigma^{j}_{ee}+\omega_s \sum_j \sigma^{j}_{ss}+\omega_g \sum_j \sigma^{j}_{gg}$, where $\omega_{e\eta}=\omega_{e}-\omega_\eta$ are the relevant atomic frequencies for the interaction with the waveguide modes.

Typically, the relaxation timescales of the reservoir are much faster than the atomic timescales. This separation of timescales justifies the so-called Born-Markov approximation that allows to calculate the evolution of the atoms, through their reduced density matrix $\rho$, after tracing out the reservoir degrees of freedom. This approximation requires that one can neglect finite propagation lengths and non Markovian effects, which requires that the maximal distance between atoms is small compared to the propagation length, $v_\mathrm{g} T$, during the time of operations, where $v_\mathrm{g}$ is the group velocity in the waveguide. In the case of a one-dimensional reservoir the evolution is then governed by the master equation \cite{gardiner_book00a}
\begin{eqnarray}
	\frac{d\rho}{dt} 
	&= {\cal L}\left[ \rho \right] 
	= - \ii \left[H_\mathrm{dd}, \rho \right] + \mathcal{L}_\mathrm{d} \left[\rho\right] \\
	H_\mathrm{dd}
	&= \sum_{m,n,\eta} \frac{\Gamma_\oned^\eta}{2} \sin \left(q_0 |z_{mn}| \right) \sigma_{e \eta}^m \sigma_{\eta e }^n \\
	\mathcal{L}_\mathrm{d} \left[\rho\right]
	&= \sum_{m,n,\eta} \frac{\Gamma_\oned^\eta}{2} \cos \left(q_0 |z_{mn}| \right) \left( \sigma_{\eta e}^m \rho \sigma_{e \eta}^n - \rho \sigma_{e \eta}^n \sigma_{\eta e}^m \right)
	+ \rm {H.c.} ,	
\end{eqnarray}
where $q_0 = q ( \omega_{eg}) = q( \omega_{es} )$ is assumed to be the same for both modes and where the renormalized spontaneous emission rate of the atoms, i.e., $\Gamma_{\oned}^\eta=\sum_q |g_q^\eta|^2 \delta(\omega_{e\eta}-\omega_{q,\eta})$, depends on both the energy dispersion ($\omega_{q,\nu}$) and guided mode profile (contained in $|g_{q}^{\eta}|$) \cite{goban13a}. Note, that the master equation above allows for two very distinct regimes for different interatomic spacings:
\begin{itemize}
	\item
	If the distance between two emitters is a multiple of $\pi / q_0 = \lambda_0/2$ the coherent terms vanish and the evolution is purely dissipative and the atoms decay through a collective operator.
	
	\item
	If the distance between two emitters is an odd multiple of $\pi / (2 q_0) = \lambda_0/4$, the dipole-dipole interactions are at their maximum.
\end{itemize}

In this work, we exploit coherent dipole-dipole interactions such that a beneficial configuration is based on two atomic mirrors (see Figure \ref{fig1}a) surrounding an atom acting as a \textit{source} for distributing atomic excitations symmetrically to the inner mirrors (\textit{target} ensembles $t_1$ and $t_2$). Collective and individual quantum jumps and other experimental imperfections cause transitions to undesired states. These transitions can be corrected by using the second guided mode and heralding measurements on the outer mirror emitters, the \textit{detector} ensembles $d_1$ and $d_2$.

In particular, the coherent dynamics in the \emph{double mirrors} configuration (in the frame rotating at the frequency of the atomic frequencies) is then described by
\begin{equation}
\label{eq:DM_coherent}
H_\mathrm{dd} 
= \frac{\Gamma_\oned^\rg}{2}  \sigma_{ge}^\src S_{eg,+}^\trg 
+ \frac{ \Gamma_\oned^\rs }{2} S_{es,-}^\dtc S_{se,-}^\trg + \mathrm{h.c.},
\end{equation}
where $S_{\alpha\beta,\pm}^{(\ry)} = S_{\alpha\beta}^{(\ry_1)} \pm  S_{\alpha\beta}^{(\ry_2)}$ are the collective operators $S_{\alpha\beta}^{(\mathrm{y})} = \sum_{j \in \mathrm{y}} \sigma_{\alpha \beta}^j$ within each ensemble, for the target ensembles $\ry=\rt$ and for the detector ensembles $\ry=\rd$. 
In addition, the Lindblad terms describing the collective decay are given by
\begin{equation}
\label{eq:SM Ld}
\mathcal{L}_{\mathrm{d}}[\rho]
	= \frac{\Gamma_\oned^\rg}{2}D_{ \sigma_{ge}^\src}[\rho]
	 +\frac{\Gamma_\oned^\rg}{2}D_{ S_{ge,-}^\trg}[\rho]
	 +\frac{\Gamma_\oned^\rs}{2}D_{ S_{se,-}^\trg}[\rho]
	 +\frac{\Gamma_\oned^\rs}{2}D_{ S_{se,+}^\dtc}[\rho],
\end{equation}
where $D_O[\rho] = 2 O \rho O^\dagger - O^\dagger O \rho - \rho O^\dagger O $.

Apart from interacting with the waveguide modes, the excited states $\ket{e}_n$ may also emit photons into free space, or even to other polarizations that do no create collective coupling between emitters. We embedded all these processes into a rate, $\Gamma^*$,  typically of the order of the natural linewidth $\Gamma_{\aa}$, which can be described through an additional Lindblad term in the Liouvillian as
\begin{equation}
{\cal L}_{*}(\rho) = 
\sum_{n,\eta} \frac{\Gamma^*}{2} \left( \sigma^n_{\eta e} \rho \sigma^n_{e \eta} - \rho \sigma^n_{e\eta} \sigma^n_{\eta e} \right)
+ \rm {h.c.}\, .
\end{equation}

Thus, one relevant figure of merit of these system is the Purcell Factor $P_{\oned}=\Gamma_{\oned}^\rg/\Gamma^*$. Another important experimental resource is the number of atoms $N$ trapped within the \textit{target} and \textit{detector} ensembles.

%-------------------------------------------------------------------------------
\subsection{Obtaining Tuneable Decay Rates by using an $M$-type structure.}

The decay rates $\Gamma_\oned^\eta$ can in principle be tuned by using an $M$-type level scheme (see Figure~\ref{fig2}(b)) which is equivalent to a $\Lambda$-type system after adiabatic elimination of the far-detuned excited states $\ket{e_\eta}$. To show this, consider the full Hamiltonian for the former system for a single atom, that is
\begin{eqnarray}
	H &= H_\aa  + H_\mathrm{b} + \sum_{\eta} \omega_{e_\eta} \sigma_{e_\eta,e_\eta}
		+ \sum_{\eta} \frac{1}{2} \left(\Omega_\eta \ee^{\ii \omega_\mathrm{L} t } \sigma_{e_\eta, e} + \mathrm{h.c.} \right) \\
		&+ \sum_{q,\eta} g_q^\eta \left(\sigma_{e_\eta,\eta} a_{q,\eta} \ee^{\ii q z} + \mathrm{h.c.} \right).
\end{eqnarray}
Transforming into the frame rotating with $H_0 = \sum_{\eta} \omega_\mathrm{L} \sigma_{e_\eta, e_\eta}$ and adiabatically eliminating the $\ket{e_\eta}$ states leads to

\begin{eqnarray}
	H_\mathrm{AE} &\approx
	\left(\omega_e - \omega_\mathrm{L} + \sum_{\eta} \frac{|\Omega_\eta|}{4 \Delta_\eta^2}\right) \sigma_{ee} 
		+ \sum_{k,\eta} \omega_k a_{q,\eta}^\dagger a_{q,\eta}  \\
		&+ \sum_{q,q',\eta} \frac{g_q^{\eta *} g_{q'}^{\eta}}{\Delta} a_{q,\eta}^\dagger a_{q',\eta} \ee^{\ii (q-q')z} 
		+ \sum_{q,\eta} \frac{g_q^\eta \Omega_\eta^*}{2 \Delta} \sigma_{e \eta} a_q \ee^{-\ii q z},
\end{eqnarray}
where $\Delta_\eta = \omega_{e_\eta} - \omega_e$. Therefore, the decay operators now turn to be $\Gamma_\oned^\eta \rightarrow \Gamma_\oned^\eta |\frac{\Omega_\eta (t)}{2 \Delta_\eta}|^2$ and can thus be controlled via the Rabi-coupling $\Omega_\eta (t)$. The Stark-shifts that are induced by this coupling can be incorporated into the corresponding frequencies.

%-------------------------------------------------------------------------------
\subsection{Simplifications}
\begin{enumerate}
	\item
	Considering that the detector is reinitialized to the state $\ket{s^\dtc}^{\otimes 2 N}$ after each step, it can only couple to the state $S_{es,-}^\dtc \ket{s^\dtc}^{\otimes 2 N}$. Both these states are dark with respect to the jump operator $S_{se,+}^\dtc$, i.e., they do not decay. Thus the decay operator for the detector ensemble can be neglected and the fourth decay term in Equation \ref{eq:SM Ld} vanishes.
	
	\item
	Due to the heralding at the end of each step, the relevant dynamics is governed by the no-jump evolution. This in turn is fully described by the non-hermitian Hamiltonian $H_\mathrm{nh} = H_\mathrm{dd} - \ii \sum_k O_k^\dagger O_k$, where $O_k$ are all the Lindblad operators of the Liouvillian of Equation \ref{eq:SM Ld}. Hence, the non-hermitian Hamiltonian is given by

\begin{eqnarray}
	H_\mathrm{nh} =&
	\frac{1}{2}\left( \Gamma_\oned^\rg  \sigma_{ge}^\src S_{eg,+}^\trg 
	+ \Gamma_\oned^\rs S_{es,-}^\dtc S_{se,-}^\trg + \mathrm{h.c.} \right) \nonumber
	\\ & - \frac{\ii}{2} \left( \Gamma_\oned^\rg \sigma_{ee}^\src 
	+ \Gamma_\oned^\rg S_{eg,-}^\trg S_{ge,-}^\trg
	+ \Gamma_\oned^\rs S_{es,-}^\trg S_{se,-}^\trg  +\Gamma^* \mathbf{1} \right).
	\label{eq:SM Hnh}
\end{eqnarray}

	\item 
	Furthermore, the total excitation number $m$ of the system is invariant under the action of the non-hermitian Hamiltonian. Hence, for superposition states $\ket{\psi} = \sum c_m \ket{\psi_m}$ each excitation can be treated separately. Therefore, we only treat dynamics of the case in which $\ket{\psi_{m-1}} \rightarrow \ket{\psi_{m}}$.
\end{enumerate}

%===============================================================================
\section{Details of the protocol in the dissipative regime.}

%-------------------------------------------------------------------------------
\subsection{Scheme of the protocol}

The protocol is based on repeated heralded additions of single (symmetric) excitations to the \emph{target} ensemble in the metastable state $\ket{s}$. After $m$ steps, the reinitialized state is $\ket{\phi_0^m} = \ket{s^\src}\otimes\ket{\psi_{m-1}}\otimes\ket{s^\dtc}^{\otimes 2N}$. An excitation is added by (see Figure \ref{fig2}c)
\begin{enumerate}
	\item 
	repumping the source atom: A single excitation is added to the system by a fast excitation of the source atom,
	\begin{equation}
	\ket{\phi_0^m} \rightarrow 
		\ket{\phi_1^m} = 
		\sigma_{es}^\src \ket{\phi_0^m} =
		\ket{e^\src} \otimes \ket{\psi_{m-1}} \otimes \ket{s^\dtc}^{\otimes 2N}.
	\end{equation}
	Clearly, this avoids any double excitations in the system.
	
	\item
	free evolution: The dipole-dipole coupling induced by the first guided mode, $\Gamma_\oned^\rg$, de-excites the source atom and distributes the excitation symmetrically over the target ensemble in the state $\ket{e}$.
	The dipole-dipole coupling induced by the second guided mode, $\Gamma_\oned^s$, moves the excitation in the target ensemble to the ground state $\ket{s}$ and excites the detector ensemble:
	\begin{equation}
	\ket{\phi^m (t)} = \ee^{-\ii H_\mathrm{nh} t} \ket{\phi_0^m},
	\end{equation}
	where $H_\mathrm{nh}$ is the non-hermitian Hamiltonian of Equation \ref{eq:SM Hnh} determining the dynamics as explained above.
	\item
	heralding on the state of the detector ensemble: After some time $T$, the detector atoms are quickly de-excited to $\ket{g}$, halting any further evolution of the system:
	\begin{equation}
	\ket{\phi_\mathrm{out}^m} = S_{ge,+}^\dtc\ket{\phi^m (T)}.
	\end{equation}
	Then, the $\ket{g}$-state of the detector ensemble is probed and if an excitation is detected, an excitation must have been added to the target ensemble and steps (i)-(iii) can be repeated after the source emitter and the detector ensemble have been reinitialized.
	
\end{enumerate}

When no excitation in the detector ensemble is detected, this can be due to several indistinguishable reasons. The case in which emission into free space or into the guided mode in either the source atom or the detector ensemble happened, don't affect the state of the target ensemble, such that one could repeat steps (i)-(iii) to try adding an excitation again. The same is true for collective emission in the $\Gamma_\oned^\rg$-mode of the target ensemble. However, spontaneous emission and collective emission in the $\Gamma_\oned^\rs$-mode of the target ensemble are not correctable without introducing additional errors. Therefore, one should restart the whole procedure if no excitation is detected. Clearly, this will lead to an exponential scaling of the success probability with the number of excitations, $m$. We discuss variations of the protocol to avoid this scaling.

\subsection{Multilevel Holstein-Primakoff Transformation and Approximation}

By using the Holstein-Primakoff-Transformation \cite{holstein40a}, it is possible to map a spin operator onto bosonic operators. This transformation is especially useful to describe the symmetric subspace of $N$ two-level quantum systems. For multilevel systems this transformation can be generalized, see e.g.~\cite{kurucz10}. In particular, one needs $d-1$ bosonic operators to describe the symmetric states of $d$-level systems.

Because this paper focuses on (effective) three-level systems we only discuss the transformation for this case here. The ground, excited and target states of one three-level system are denoted by $\ket{g}$, $\ket{e}$ and $\ket{s}$. Any symmetric state can then be described by 
\begin{equation}
\ket{m_s, m_e} 
\propto \mathrm{sym}
\left( 
\ket{s}^{\otimes m_s} \otimes \ket{e}^{\otimes m_e} \otimes \ket{g}^{\otimes N-m_s-m_e}
\right).
\end{equation}
We denote the two bosonic operators by $b_s$ for the annihilation operator of $\ket{s}$-state, and $b_e$ for the excited state. These operators should satisfy
\begin{eqnarray}
	b_s \ket{m_s, m_e} = \sqrt{m_s} \ket{m_s-1, m_e},  \ \hps^\dagger \hps \ket{m_s, m_e} = m_s \ket{m_s, m_e},\\
	b_e \ket{m_s, m_e} = \sqrt{m_e} \ket{m_s, m_e-1},  \ \hpe^\dagger \hpe \ket{m_s, m_e} = m_e \ket{m_s, m_e}\,,
\end{eqnarray}
and hence, they also commute $[\hps, \hpe] = [\hps^\dagger, \hpe^\dagger] = [\hps, \hpe^\dagger] = 0$.

The spin operators $S_{\alpha \beta} = \sum_{j=1}^{N} \sigma_{\alpha \beta}^j$ can then be expressed by the above bosonic operators as 
\begin{eqnarray}
	S_{gg} = N - s, \
	&S_{ss} = \hps^\dagger \hps,\
	&S_{ee} = \hpe^\dagger \hpe,\\
	S_{sg} = \hps^\dagger \sqrt{N-s},\
	&S_{eg} = \hpe^\dagger \sqrt{N-s},\
	&S_{se} = \hps^\dagger \hpe,
\end{eqnarray}
where $s = \hps^\dagger \hps + \hpe^\dagger \hpe$. In the low excitations regime, $m= \langle s \rangle \ll N$, the operators are linear in each bosonic operator up to first order $\mathcal{O}\left(\frac{m}{N} \right)$.

\subsection{Dynamics Within Holstein-Primakoff-Approximation: Probabilities}

The spin operators of the target ensemble can be mapped to commuting bosonic operators $b_{\eta,j}$ in the low excitation regime $m \ll N$, where $m$ is the number of excitations (in either $\ket{s}$ or $\ket{e}$) of the state. The spin operators can then be replaced by
\begin{eqnarray}
	S_{gg,1(2)}^\trg &= N - \hpsi{1(2)}^\dagger \hpsi{1(2)} - \hpei{1(2)}^\dagger \hpei{1(2)} \approx N,\\
	S_{eg,1(2)}^\trg &= \hpei{1(2)}^\dagger \sqrt{N - \hpsi{1(2)}^\dagger \hpsi{1(2)} - \hpei{1(2)}^\dagger \hpei{1(2)}} \approx \sqrt{N} \hpei{1(2)}^\dagger, \\
	S_{sg,1(2)}^\trg &= \hpsi{1(2)}^\dagger \sqrt{N - \hpsi{1(2)}^\dagger \hpsi{1(2)} - \hpei{1(2)}^\dagger \hpei{1(2)}} \approx \sqrt{N} \hpsi{1(2)}^\dagger, \\
	S_{se,1(2)}^\trg &= \hpsi{1(2)}^\dagger \hpei{1(2)}.
		\label{eq:HPT}
\end{eqnarray}

Within this approximation, the relevant states coupled trough the non-hermitian Hamiltonian of Equation \ref{eq:SM Hnh} are
\begin{eqnarray}
	\ket{\phi_1^m} &= \sigma_{es}^\src \ket{\phi_0^m},\\
	\ket{\phi_2^m} &= 
		\sigma_{gs}^\src \frac{1}{\sqrt{2}} \left( \hpei{1}^\dagger + \hpei{2}^\dagger \right) \ket{\phi_0^m},\\
	\ket{\phi_3^m} &= 
		\sigma_{gs}^\src \frac{1}{\sqrt{2m}} \left( \hpsi{1}^\dagger - \hpsi{2}^\dagger \right) \frac{1}{\sqrt{2 N}} S_{es,-}^\dtc \ket{\phi_0^m} \propto
		\sigma_{gs}^\src S_{es,-}^\dtc \ket{\phi_0^{m+1}} ,
	\label{eq:basisHP}
\end{eqnarray}
where the reference state is $\ket{\phi_0^m} = \ket{s^\src} \otimes \ket{\psi_{m-1}} \otimes \ket{s^\dtc}^{\otimes 2N}$.
The normalization of $\ket{\phi_3^m}$ is due to the fact, that we consider an initial state of the target ensemble $\ket{\psi_{m-1}}$, which already contains $m$ excitations. In particular 
\begin{equation}
\ket{\psi_m}=\frac{1}{\sqrt{2^m \cdot m!}} \left(\hpei{1}^\dagger - \hpei{2}^\dagger \right)^m \ket{g^\trg}^{2N}.
\end{equation}

In the basis of the states $\ket{\phi_i^m}$, the non-hermitian Hamiltonian in the $m$-excitation subspace can written as
\begin{equation}
\widetilde{H}_\mathrm{nh} 
= \frac{1}{2} \left(
\begin{array}{ccc}
-\ii \left(\Gamma_\oned^\rg + \Gamma^* \right) & \sqrt{2N} \Gamma_\oned^\rg & 0\\
\sqrt{2N} \Gamma_\oned^\rg & -\ii \left(m \Gamma_\oned^\rs + \Gamma^*\right) &  \sqrt{2N m} \Gamma_\oned^\rs\\
0 & \sqrt{2N m} \Gamma_\oned^\rs & -\ii \Gamma^*\\
\end{array}
\right).
\end{equation}

A maximal population transfer between $\ket{\phi_1^m}$ and $\ket{\phi_3^m}$ is reached for tunable $ \Gamma_\oned^\rg = \sqrt{m} \Gamma_\oned^\rs$ after time $T = \sqrt{2} \pi \left( \sqrt{2N} \Gamma_\oned^\rg \right)^{-1}$. Other choices for the parameters, in particular ones allowing for $\Gamma_\oned^\rg = \Gamma_\oned^\rs$, may also lead to high enough success probability and are discussed later.

The protocol is limited by the success probability for each of the steps $\ket{\psi_{m-1}} \rightarrow \ket{\psi_{m}}$, that is, the probability of one of the detector atoms being in $\ket{g}$ and therefore the generation of $\ket{\psi_m}$ in the target ensemble. To quantify it, one needs to calculate 
\begin{equation}
p_{m-1\rightarrow m} = \|S_{ge,+}^\dtc \ee^{-\ii H_\mathrm{nh} T} \ket{\phi_1^m} \|^2,
\label{eq:p_succ}
\end{equation}
after the de-excitation with $S_{ge,+}^\dtc$. In the optimal case, the success probability of the heralding measurement (Figure \ref{fig:pSucc}a) is 
\begin{equation}
p_{m-1\rightarrow m} \approx 
\mathrm{exp} 
\left[ 
- \frac{\sqrt{2} \pi}{8 \sqrt{2N}} \left( 3 + 2 \sqrt{m} + 8 P_\oned^{-1} \right)
\right].
\label{eq:SM pm}
\end{equation}

The scaling originates in the fact that the process is very fast, $T\propto N^{-1/2}$, and that the non-hermitian terms scale as $\Gamma_\oned^\rg$ and $m \Gamma_\oned^\rs = \sqrt{m} \Gamma_\oned^\rg$. The scalings arise from the population of the specific states which are subject to the quantum jumps.

Clearly, the fidelity of the process, that is the overlap between the goal state $\ket{\phi_3^m}$ and the final projected state $S_{ge,+}^\dtc \ee^{-\ii H_\mathrm{nh} T} \ket{\phi_1^m}$ after normalization is unity within the Holstein-Primakoff approximation. We neglect errors originating from finite detection efficiencies and dark counts, because the detection via the detector ensemble can be repeated as many times as necessary.

By repeatedly adding excitations, we can accumulate several excitations within the same mode $\ket{\psi_m} \propto \left(S_{sg,-}^{\trg}\right)^m \ket{\psi_0}$. Clearly, the average number of operations  $R_m = \prod_{k=1}^{m} p_{k-1\rightarrow k}^{-1}$ to obtain this state is, according to the previous discussion, exponential in the number of excitations and scales approximately as $R_m\propto e^{m\sqrt{\frac{m}{N}}}$ for large excitation numbers $1\ll m\ll N$.

Superpositions are obtained by alternately adding a single excitation and applying a displacement operator \cite{dakna99a}. The displacement operators can be easily applied through a well-controlled microwave transition between the metastable ground states of the target atoms.

%-------------------------------------------------------------------------------
\subsection{Beyond Holstein-Primakoff Approximation: fidelities}

Without the Holstein-Primakoff-Approximation the non-hermitian Hamiltonian couples to additional states. In particular, the decay operator $\Gamma_\oned^\rg S_{eg,-}^\trg S_{ge,-}^\trg$ couples the state $S^\trg_{eg,+} \ket{\phi_1^m}$ to states that are linearly independent of the basis states defined in Equations \ref{eq:basisHP}. Because the operators no longer satisfy the bosonic commutation relations, a coupling between the symmetric and antisymmetric states of the target ensemble is possible. In fact, instead of three orthonormal states one now has to consider $4m+1$ orthonormal states. The states under consideration are always symmetric in each ensemble and thus a target ensemble state can be denoted by
\begin{equation}
\ket{k_1, l_1; k_2, l_2} \propto 
S_{sg,1}^{k_1} S_{eg,1}^{l_1} S_{sg,2}^{k_2} S_{eg,2}^{l_2} \ket{g}^{\otimes 2N},
\end{equation}
where for simplicity the superindex $\trg$ was omitted. The restrictions on $(k_1, l_1, k_2, l_2)$, i.e., $k_1+l_1+k_2+l_2 = m-1$ or $m$ and $(l_1, l_2) = (0,0), (1,0), (0,1)$, leave the $4m+1$ states,
\begin{eqnarray}
	\ket{e^\src} \otimes &\ket{m-1-i, 0; i, 0} \otimes \ket{s^\dtc}^{ \otimes 2N},\ \ i=0, \ldots m-1, \\
	\ket{g^\src} \otimes &\ket{m-1-i, 1; i, 0} \otimes \ket{s^\dtc}^{ \otimes 2N},\ \ i=0, \ldots m-1, \\
	\ket{g^\src} \otimes &\ket{m-1-i, 0; i, 1} \otimes \ket{s^\dtc}^{ \otimes 2N},\ \ i=0, \ldots m-1, \\	
	\ket{g^\src} \otimes &\ket{m-i, 0; i, 0} \otimes S_{es,-} \ket{s^\dtc}^{ \otimes 2N}, \ i=0, \ldots m.
\end{eqnarray}

The multitude of additional states that $H_\mathrm{nh}$ couples to may lead to a non-unit overlap with the goal target state $\ket{\psi_m^\mathrm{goal}} \propto S_{sg,-}^m \ket{g^\trg}^{\otimes 2N}$. Clearly, in the limit of large ensemble size $N$, the fidelity has to go to unity to agree with the Holstein-Primakoff-Approximation.

The numerical results can then be obtained by applying the exact Holstein-Primakoff-Transformation, see Equation \ref{eq:HPT} and using a cut-off parameter of $m+1$ for the operators $\hpsi{i}$ and $2$ for the operators $\hpei{i}$. For the generation of $m$ excitations, the new input state has to be obtained from the output state of the step before, i.e.
\begin{eqnarray}
	\ket{\phi_\mathrm{in}^m} &= \ket{e^\src} \otimes \ket{\psi_\mathrm{in}^m} \otimes \ket{s^\dtc}^{\otimes 2N}, \\
	\ket{\psi_\mathrm{in}^m} &\propto \Tr_\mathrm{s,d} \left( S_{ge,+}^\dtc \ee^{-\ii H_\mathrm{nh} T} \ket{\phi_\mathrm{in}^{m-1}} \right).
\end{eqnarray}

The results from the full numerical analysis (Figure \ref{fig:pSucc}) agree very well with the results obtained by applying the Holstein-Primakoff-Approximation. The infidelity scales as $I_m \approx 0.061 \frac{m^´2}{N^2}$ for $1 \ll m \ll N$, where the prefactor was found by a numerical fit. The fidelity is independent of the Purcell Factor because the error stems from non-linear corrections to the Holstein-Primakoff picture that enter through collective rather than spontaneous emission events which affect every state in the the same way.

% ------------------------------------------------------------------------------
\subsection{Variations of the protocol}

\begin{figure}[t]
	\centering
	\includegraphics[width=0.95\textwidth]{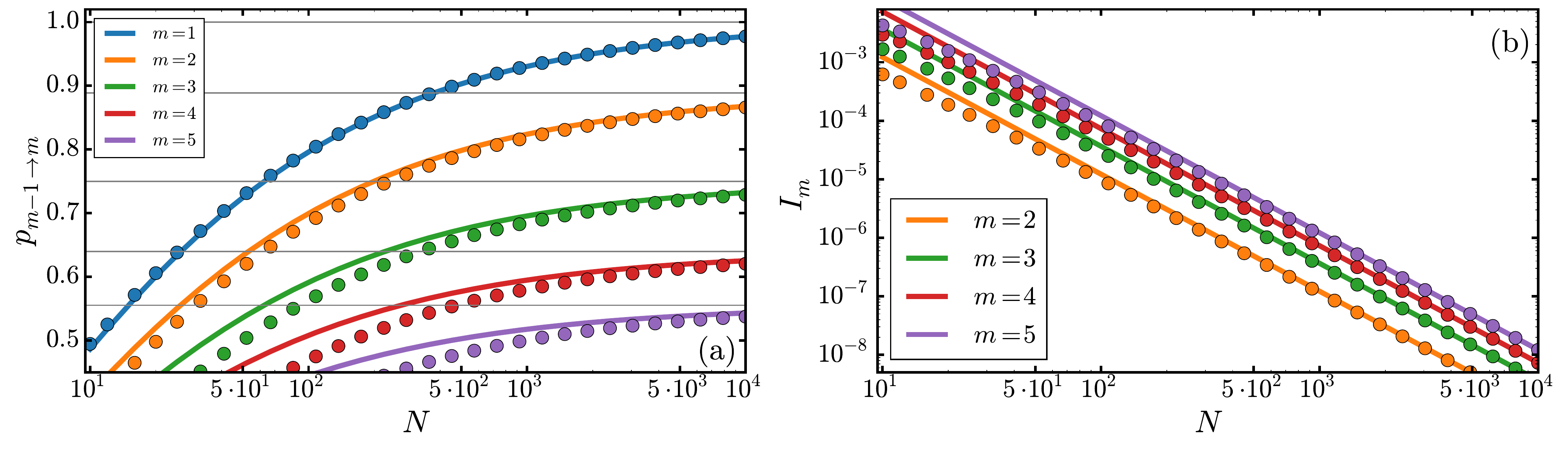}
	\caption{(a) For the success probability, Equation \ref{eq:SM pm2} is a good approximation (solid line) of the full solution (circles). The gray horizontal lines depict the limit of the success probability, Equation \ref{eq:SM pm2_limit}, caused by an incomplete population transfer. The results here do not include the $P_\oned$ term.
	(b) The overlap with the goal state is $1$ within the Holstein-Primakoff-Approximation and for $m=0$ and slightly deviates from unity by $1/N^2$ for the full solution. As expected, they agree in the limit of large ensemble sizes $N \gg 1$.}
	\label{fig:pSucc2}
\end{figure}

The scheme previously proposed can be modified if some of the demanded ingredients are not available. The goal is to maximize the success probability (Equation \ref{eq:p_succ}) constrained by the parameters that are experimentally achievable:

\begin{enumerate}
	\item \emph{Fixed ratio $\Gamma_\oned^\rs / \Gamma_\oned^\rg$}.
	The previous analysis showed that, e.g., when $\Gamma_\oned^\rs / \Gamma_\oned^\rg$ can be tuned around orders of $\sqrt{m} = \mathcal{O}(1)$, then almost unit probability can be reached.
	If on the contrary the ratio between decay rates is fixed, e.g., $\Gamma_\oned^\rs = \Gamma_\oned^\rg$, then the maximal success probability occurs at $T = \frac{2 \pi}{\sqrt{2N(m+1)}\Gamma_\oned^\rs}$ and scales within the Holstein-Primakoff Approximation as
	\begin{equation}
	\tilde{p}_{m-1\rightarrow m}=\frac{4 m}{(m+1)^2}
	\rme^{\left[-\frac{2 \pi}{\sqrt{2 N (m+1)}} 
	\left( \frac{3 m^2+m+1}{2 (m+1)^2} + P_\oned^{-1} \right)\right]}.
	\label{eq:SM pm2}
	\end{equation}
	The success probability goes to a constant value for $m>1$:
	\begin{equation}
	\tilde{p}_{m-1\rightarrow m} \stackrel{N\rightarrow \infty}{\longrightarrow} 
	\frac{4 m}{(m+2)^2}.
	\label{eq:SM pm2_limit}
	\end{equation}
	whereas the resulting infidelity still scales as
	\begin{equation}
		I_{m} \approx 0.061 \frac{m(m-1)}{N^2}
	\end{equation}
	for $m \ll N$, are depicted in Figure \ref{fig:pSucc2}. The prefactor was obtained from a numerical fit.
	
	\item \emph{Replacing fast $\pi$-pulse for continuous driving}.
	The fast $\pi$-pulse on the source atom at the beginning and on the detector ensemble at the end of each step can be avoided by applying a continuous external field of finite Rabi coupling strength to the respective transitions, i.e.,  the $\ket{s}-\ket{e}$-transition of the source atom and the $\ket{g}-\ket{e}$-transition of the detector ensemble. The dynamics (see Figure \ref{fig:Variation_Om}a) then contains five states (within the Holstein-Primakoff-Approximation). These states are $\ket{\phi_m}$, for $m=0,..3$ from the definitions above and $\ket{\phi_4} \propto S_{ge}^\dtc \ket{\phi_3}$. A full population transfer to the desired state is obtained for $\Gamma_\oned^\rg = \sqrt{m} \Gamma_\oned^\rs$ as for the main protocol and for a coupling strength of $\Omega = \sqrt{\frac{2}{3}} \sqrt{2N} \Gamma_\oned^\rg$.
	The success probability is maximized at time $T = \pi \sqrt{6} \left(\sqrt{2N} \Gamma_\oned^\rg\right)^{-1} = 3 \pi / \Omega$ and in this optimal case the scaling of the success probability with $N$, $m$ and $P_\oned$ is then approximately (see Figure \ref{fig:Variation_Om}b)
	\begin{equation}
		\widetilde{p}_{m-1\rightarrow m} = \mathrm{exp}
		\left[-\frac{\sqrt{6} \pi}{\sqrt{2 N}} 
		\left( \frac{10 + 9 \sqrt{m}}{64} + \frac{29}{64} P_\oned^{-1} \right)\right].
		\label{eq:SM q_m}
	\end{equation}
	That means, the scaling remains the same and only some prefactors in the exponent change slightly.
	
	\begin{figure}[t]
		\centering
		\includegraphics[width=0.95\textwidth]{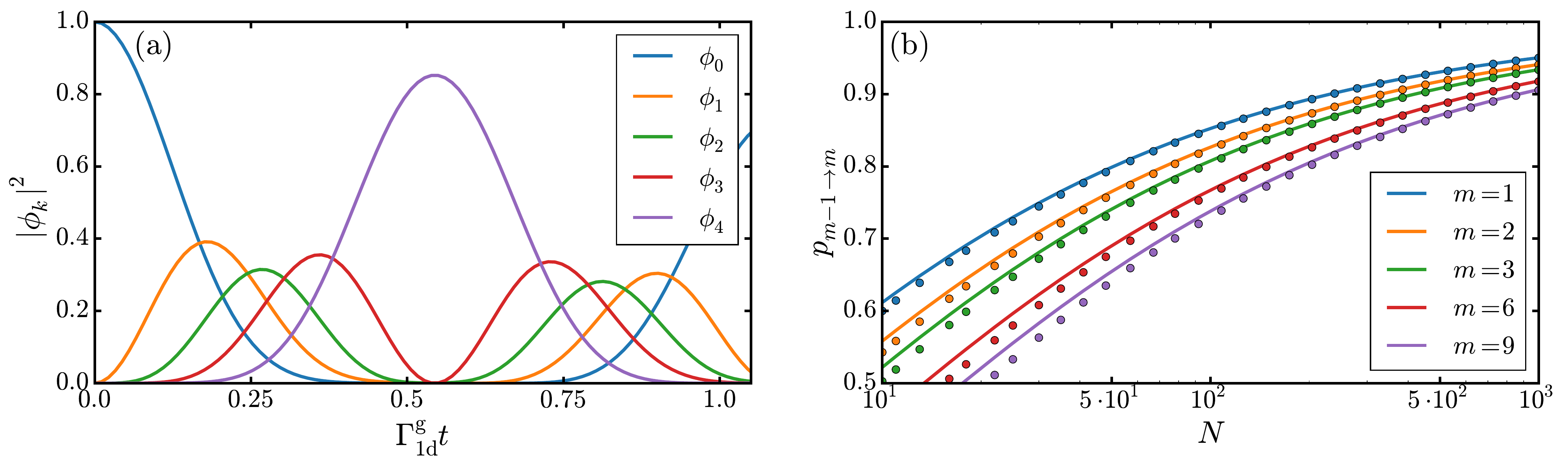}
		\caption{(a) The time evolution when including a finite Rabi coupling strength and when one needs to consider additional states. For $N=100$, $m=1$ and the optimal parameters discussed in the text a full population transfer (when quantum jumps are neglected) is still possible. (b) For the optimal parameters and the time for a full population transfer $T$, the success probability of the Holstein-Primakoff-approximated evolution scales as in Equation \ref{eq:SM q_m} (solid lines) agrees well with the numerical results obtained by the Holstein-Primakoff-Transformation (circles).}
		\label{fig:Variation_Om}
	\end{figure}
	
	\item \emph{Using a single guided mode.} 
	Even if only a single guided mode is available, say the $\Gamma_\oned^\rg$-mode, the proposed protocol can still be applied. To ensure that the fidelity of the desired state is still close to 1, one requires an additional metastable state $\ket{c}$ in the target ensemble to which spontaneous emission, $\Gamma^*_c$, is strongly suppressed, and good control over a $\pi$-pulse between the two metastable ground states $\ket{g}$ and $\ket{s}$.
	
	The step (ii) is then split up into two steps as schematically depicted in Figure \ref{fig:Variations}a. The first one involves the transfer of excitations from the source atom to the target ensemble. The excitation is stored in the additional metastable state $\ket{c}$. The role of the $\ket{g}$ and $\ket{s}$ state is then reversed by applying a well controlled $\pi$-pulse between these states on every atom. Finally, a $\pi$-pulse from $\ket{c}$ to $\ket{e}$ is applied and the dipole coupling transfers the excitation to the detector ensemble. Clearly, the detector atoms should be in the $\ket{g}$-state initially for this and the measurement is then done on the $\ket{s}$-state, to which the detector ensemble should be de-excited to.
	
	The infidelity then contains additional terms $ N \Delta(\Omega_{MW} T)^2$ and $\frac{\Gamma^*_c}{\sqrt{N} \Gamma_\oned^\rg}$,	where optimally the pulse area is $\Omega_{MW} T = \pi$ for the full population transfer between the metastable ground states and $\Delta(\Omega_{MW} T)$ is the deviation from this value. As before, the error induced in every step will accumulate such that the final infidelity $I_m = \sum_{k=1}^{m} I_{k-1\rightarrow k}$.
	
	\item \emph{Adding $m$ excitations at once.} Instead of generating single excitations in every step through a single source atom, one could in principle also use a source ensemble of size $m$ and transfer all excitations to the target ensemble to generate $m$ collective excitations at once. However, the source atoms are then superradiant and decay with an enhanced decay rate of at least $m \Gamma_\oned$. On the other hand, the dipole couplings are only enhanced by $\sqrt{m N }\Gamma_\oned$, which implies that the probability would still scale exponentially in $\sqrt{\frac{m}{N}}$. In addition, one requires a measurement device which can resolve the excitation number of the detector ensemble to guarantee the transfer of $m$ excitations to the target ensemble. Even if that is possible, e.g., the probability for generating two excitations at once, $p_{0\rightarrow2}$, is lower than the probability, $p_{0\rightarrow1} p_{1\rightarrow2}$, obtained through the original protocol (see Figure \ref{fig:Variations}(d)).
\end{enumerate}

\begin{figure}[t]
	\centering
	\includegraphics[width=0.60\textwidth]{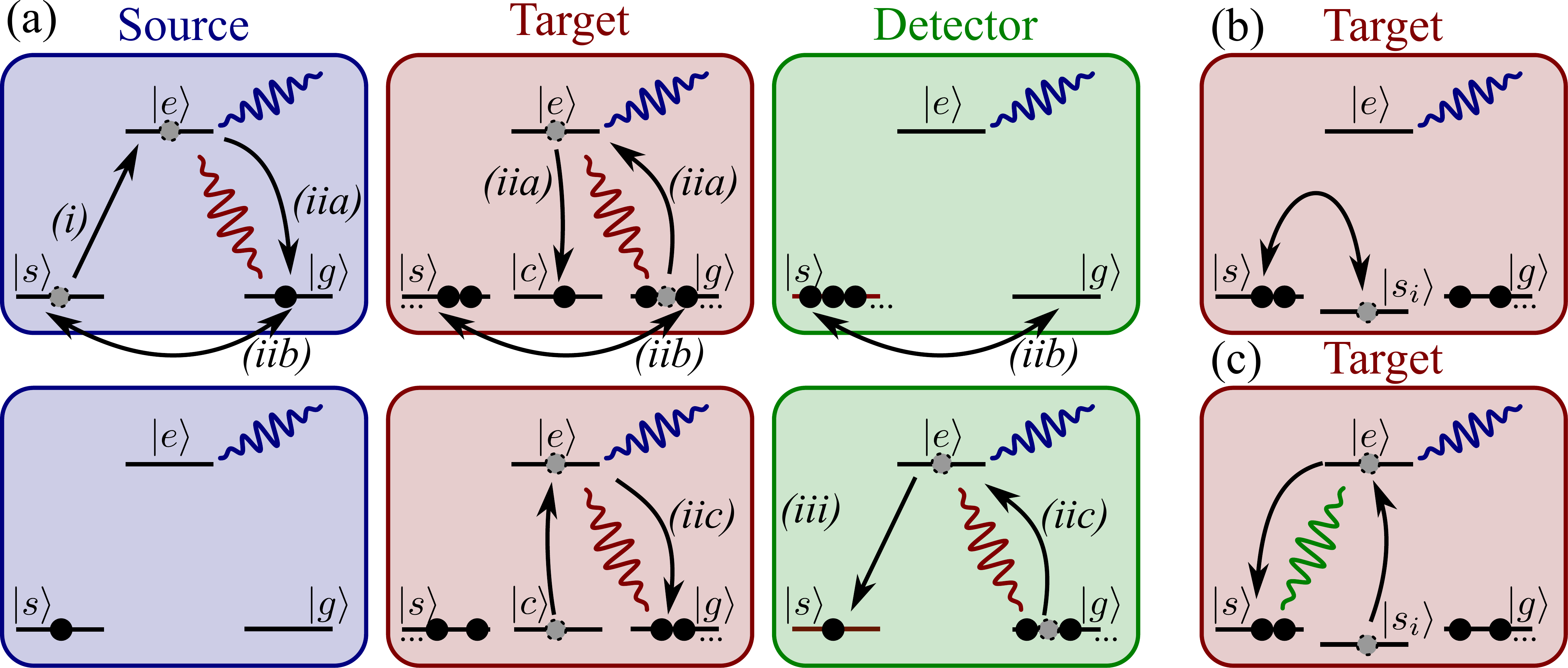}
	\includegraphics[width=0.37\textwidth]{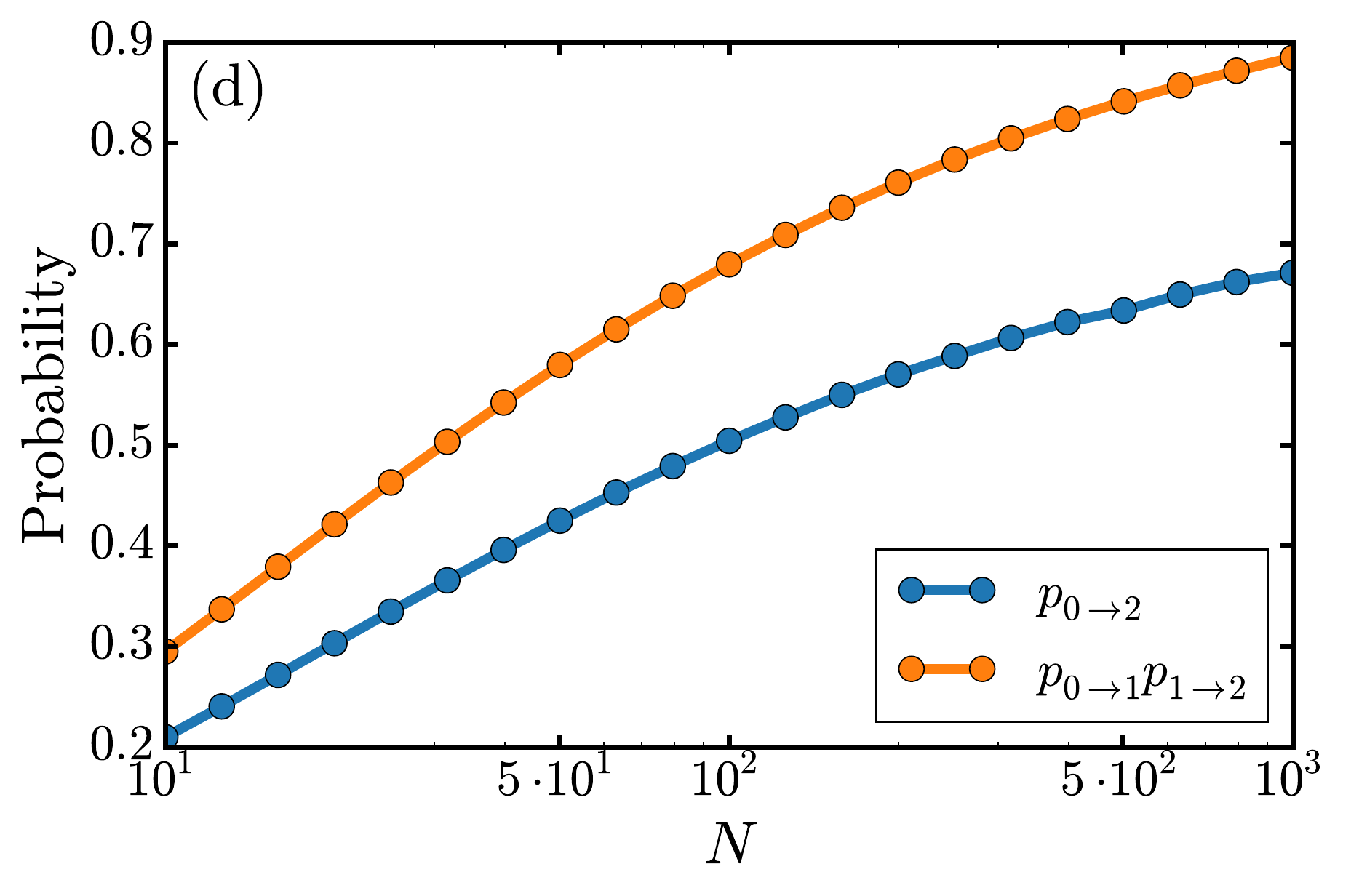}
	\caption{a) If only a single guided mode is available, the protocol can still be applied if another metastable state with strongly suppressed spontaneous emission is available. b) Beam-splitter-like transformations between metastable states are obtained by applying a corresponding external driving field. c) Excitations can also be added by using the decay through the guided mode. d) The heralding probability for generating two excitations at once is lower than the heralding probability of the original protocol.}
	\label{fig:Variations}
\end{figure}

%-------------------------------------------------------------------------------
\subsection{Intermediate Storage in further metastable ground states}

For a large excitation number $m$ the $\sqrt{m/N}$-term in the success probability of Equation \ref{eq:SM pm} causes a fast decay of the probability with $m$. This is due to the fact that the target ensemble decays super-radiantly with respect to the $\Gamma_\oned^\rs$-mode, that is with an enhanced decay rate of $m \Gamma_\oned^\rs$. This can be avoided if the $\ket{s}$-state of the target ensemble does not contain any excitations in the beginning of each step. The excitations are then not accumulated in the $\ket{s}$-state, but in other metastable ground states of the target ensemble, $\ket{s_i}$. The number of additional metastable ground states necessary depends on the approach of how to add up the excitations a posteriori. 

The success probability of adding another excitations when the $m$ excitations are stored in different hyperfine levels, $\ket{s_i}$, different from $s$ is modified as follows: 
\begin{equation}
p_{m-1\rightarrow m} \approx
\mathrm{exp} 
\left[ 
- \frac{\sqrt{2} \pi}{8 \sqrt{2N}} \left( 5 + 8 P_\oned^{-1} \right)\,,
\right].
\end{equation}
which, as expected, does not show the additional scaling with $\sqrt{m}$.

Moreover, it is important to highlight another difference with respect to the previous protocol. Previously, when a heralding measurement failed we reinitialize the target ensemble back to $\ket{g}$ destroying the stored $m$ excitations. However, in this case the excitations stored in $\ket{s_i}$ can be salvaged with minimal error by using an appropriate repumping scheme in which the symmetry of the state is unaffected. This is achieved by switching $\Gamma_\oned^\rs=0$, applying a repumping Hamiltonian $H_\mathrm{RP} = \Omega S_{se,+}^\trg + \mathrm{h.c.}$ and waiting for this state to decay to the state $\ket{g}$. In fact, this decay is superradiant with respect to the $\Gamma_\oned^\rg$-mode due to the symmetry of the state of the target ensemble. There are two types of errors: i) a spontaneous jump in the target ensemble can already have occured before the repumping step, in which case the error probabiliy scales as $\frac{1}{\sqrt{N} P_\oned}$; or ii) a collective jump in the target ensemble has occured (which happens with probability $\frac{1}{\sqrt{N}}$), but during the repumping step a spontaneous jump happend, in which case the total error probability scales as $\frac{1}{\sqrt{N}} \frac{1}{N P_\oned}$. Because the overlap with the symmetric state scales as $1-\frac{m}{N}$ after spontaneous emission events, the error is upper bounded by $\frac{\Gamma^*}{N \sqrt{N} \Gamma_\oned^\rg}$ after the repumping step.

Concerning the methods to accumulate excitations into a single level, there are multiple approaches that sketch briefly:
\begin{enumerate}
	\item
	Beam-splitter-like transformations (see Figure \ref{fig:Variations}b) can be used to accumulate excitations, for a detailed analysis see for example \cite{gonzaleztudela16a}. A ``beam-splitter'' between the metastable ground states $\ket{s_i}$ can be applied by switching on an external field coupling two of those states for a certain time. The pulse area determines the transmittance of the beam-splitter. To add up the population in two of these states, one has to herald on detecting no excitations in one of the metastable states. If the excitations are added one-by-one, the final success probability is exponential in the number of excitations $m$. However, when doubling the excitations at each step, one can reach polynomial scaling for Fock states and subexponential scaling for superpositions.
	
	\item
	In principle, excitations in the metastable states $\ket{s_i}$ can also be added by using the $\Gamma_\oned^s$-mode of the waveguide and the detector ensemble. This case is clearly equivalent to adding the excitations directly one by one. In principle, this scheme can be used to add multiple excitations at once. These are however difficult to detect and require number-resolving measurements in the detector ensemble.
	
	\item
	To increase the success probability significantly, one can use the fact that an excited state has to decay after some time, either through the guided mode or spontaneously to free space (or other guided modes different to the mode of interest). If only decay to the level $\ket{s}$ is possible (see Figure \ref{fig:Variations}c) and the detector atom is decoupled, an excitation in $\ket{e}$ is certainly added to $\ket{s}$. However, because no heralding measurements are applied, the fidelity is reduced due to spontaneous emission events. For adding excitations one by one, the fidelity is reduced by $\frac{\Gamma^*}{m \Gamma_\oned^\rs}$ in the step $m-1\rightarrow m$. If only excitations are added one-by-one, the averaged infidelity at the end is $\sum_{k=1}^m \frac{\Gamma^*}{k \Gamma_\oned^\rs}$. In principle, this can be improved, e.g., by adding two or even more excitations at once.
\end{enumerate}

%===============================================================================
\section*{References}
\bibliographystyle{unsrt}

\bibliography{Ref_DM}

\begin{thebibliography}{10}

\bibitem{migdall_book13}
Alan Migdall, Sergey~V. Polyakov, Jingyun Fan, and C.~Joshua.
\newblock {\em {Experimental Methods in the Physical Sciences Experimental
  Methods in the Physical Sciences, Volume 45, Single-Photon Generation and
  Detection Physics and Applications}}.
\newblock Science Direct, Academic Press, 2013 Maryland Heights, MO United
  State, 2013.

\bibitem{kurtsiefer00}
Christian Kurtsiefer, Sonja Mayer, Patrick Zarda, and Harald Weinfurter.
\newblock Stable solid-state source of single photons.
\newblock {\em Phys. Rev. Lett.}, 85(2):290--293, jul 2000.

\bibitem{zwiller03}
V.~Zwiller, T.~Aichele, W.~Seifert, J.~Persson, and O.~Benson.
\newblock Generating visible single photons on demand with single {InP} quantum
  dots.
\newblock {\em Appl. Phys. Lett.}, 82(10):1509, 2003.

\bibitem{pechal14}
M.~Pechal, L.~Huthmacher, C.~Eichler, S.~Zeytino{\u{g}}lu, A.~A. Abdumalikov,
  S.~Berger, A.~Wallraff, and S.~Filipp.
\newblock Microwave-controlled generation of shaped single photons in circuit
  quantum electrodynamics.
\newblock {\em Phys. Rev. X}, 4(4), oct 2014.

\bibitem{law97}
C.~K. Law and H.~J. Kimble.
\newblock Deterministic generation of a bit-stream of single-photon pulses.
\newblock {\em Journal of Modern Optics}, 44(11-12):2067--2074, nov 1997.

\bibitem{keller04}
Matthias Keller, Birgit Lange, Kazuhiro Hayasaka, Wolfgang Lange, and Herbert
  Walther.
\newblock Continuous generation of single photons with controlled waveform in
  an ion-trap cavity system.
\newblock {\em Nature}, 431(7012):1075--1078, oct 2004.

\bibitem{beugnon06}
J.~Beugnon, M.~P.~A. Jones, J.~Dingjan, B.~Darqui{\'{e}}, G.~Messin,
  A.~Browaeys, and P.~Grangier.
\newblock Quantum interference between two single photons emitted by
  independently trapped atoms.
\newblock {\em Nature}, 440(7085):779--782, apr 2006.

\bibitem{thompson06}
J.~K. Thompson.
\newblock A high-brightness source of narrowband, identical-photon pairs.
\newblock {\em Science}, 313(5783):74--77, jul 2006.

\bibitem{barros09}
H~G Barros, A~Stute, T~E Northup, C~Russo, P~O Schmidt, and R~Blatt.
\newblock Deterministic single-photon source from a single ion.
\newblock {\em New Journal of Physics}, 11(10):103004, oct 2009.

\bibitem{muecke13}
Martin M{\"u}cke, Joerg Bochmann, Carolin Hahn, Andreas Neuzner, Christian
  N{\"o}lleke, Andreas Reiserer, Gerhard Rempe, and Stephan Ritter.
\newblock Generation of single photons from an atom-cavity system.
\newblock {\em Phys. Rev. A}, 87(6):063805, jun 2013.

\bibitem{chou04}
C.~W. Chou, S.~V. Polyakov, A.~Kuzmich, and H.~J. Kimble.
\newblock Single-photon generation from stored excitation in an atomic
  ensemble.
\newblock {\em Phys. Rev. Lett.}, 92(21):213601, may 2004.

\bibitem{chen06}
Shuai Chen, Yu-Ao Chen, Thorsten Strassel, Zhen-Sheng Yuan, Bo~Zhao, Jörg
  Schmiedmayer, and Jian-Wei Pan.
\newblock Deterministic and storable single-photon source based on a quantum
  memory.
\newblock {\em Phys. Rev. Lett.}, 97(17):173004, oct 2006.

\bibitem{simon07}
Jonathan Simon, Haruka Tanji, James~K. Thompson, and Vladan Vuleti{\'{c}}.
\newblock Interfacing collective atomic excitations and single photons.
\newblock {\em Phys. Rev. Lett.}, 98(18):183601, may 2007.

\bibitem{eisaman04}
M.~D. Eisaman, L.~Childress, A.~Andr{\'{e}}, F.~Massou, A.~S. Zibrov, and M.~D.
  Lukin.
\newblock Shaping quantum pulses of light via coherent atomic memory.
\newblock {\em Phys. Rev. Lett.}, 93(23):233602, nov 2004.

\bibitem{ourjoumtsev06}
A.~Ourjoumtsev.
\newblock Generating optical schrodinger kittens for quantum information
  processing.
\newblock {\em Science}, 312(5770):83--86, apr 2006.

\bibitem{bimbard10}
Erwan Bimbard, Nitin Jain, Andrew MacRae, and A.~I. Lvovsky.
\newblock Quantum-optical state engineering up to the two-photon level.
\newblock {\em Nature Photonics}, 4(4):243--247, feb 2010.

\bibitem{lang13}
C~Lang, C~Eichler, L~Steffen, JM~Fink, MJ~Woolley, A~Blais, and A~Wallraff.
\newblock Correlations, indistinguishability and entanglement in hong-ou-mandel
  experiments at microwave frequencies.
\newblock {\em {Nat. Phys.}}, 9(6):345--348, 2013.

\bibitem{cooper13}
Merlin Cooper, Laura~J. Wright, Christoph Söller, and Brian~J. Smith.
\newblock Experimental generation of multi-photon fock states.
\newblock {\em Opt. Express}, 21(5):5309, feb 2013.

\bibitem{dellanno06}
Fabio Dell'Anno, Silvio~De Siena, and Fabrizio Illuminati.
\newblock Multiphoton quantum optics and quantum state engineering.
\newblock {\em Physics Reports}, 428(2-3):53--168, may 2006.

\bibitem{laucht12}
A.~Laucht, S.~Pütz, T.~Günthner, N.~Hauke, R.~Saive,
  S.~Fr{\'{e}}d{\'{e}}rick, M.~Bichler, M.-C. Amann, A.~W. Holleitner,
  M.~Kaniber, and J.~J. Finley.
\newblock A waveguide-coupled on-chip single-photon source.
\newblock {\em Phys. Rev. X}, 2(1):011014, mar 2012.

\bibitem{lodahl15}
Peter Lodahl, Sahand Mahmoodian, and S{\o}ren Stobbe.
\newblock Interfacing single photons and single quantum dots with photonic
  nanostructures.
\newblock {\em Reviews of Modern Physics}, 87(2):347--400, may 2015.

\bibitem{yu14a}
S-P Yu, JD~Hood, JA~Muniz, MJ~Martin, Richard Norte, C-L Hung, Se{\'a}n~M
  Meenehan, Justin~D Cohen, Oskar Painter, and HJ~Kimble.
\newblock Nanowire photonic crystal waveguides for single-atom trapping and
  strong light-matter interactions.
\newblock {\em Appl. Phys. Lett.}, 104(11):111103, 2014.

\bibitem{thompson13a}
J.~D. Thompson, T.~G. Tiecke, N.~P. de~Leon, J.~Feist, A.~V. Akimov,
  M.~Gullans, A.~S. Zibrov, V.~Vuletic, and M.~D. Lukin.
\newblock Coupling a single trapped atom to a nanoscale optical cavity.
\newblock {\em Science}, 340(6137):1202--1205, 2013.

\bibitem{tiecke14a}
TG~Tiecke, JD~Thompson, NP~de~Leon, LR~Liu, V~Vuleti{\'c}, and MD~Lukin.
\newblock Nanophotonic quantum phase switch with a single atom.
\newblock {\em Nature}, 508(7495):241--244, 2014.

\bibitem{goban15a}
A~Goban, C-L Hung, JD~Hood, S-P Yu, JA~Muniz, O~Painter, and HJ~Kimble.
\newblock Superradiance for atoms trapped along a photonic crystal waveguide.
\newblock {\em Phys. Rev. Lett.}, 115(6):063601, aug 2015.

\bibitem{vetsch10a}
E~Vetsch, D~Reitz, G~Sagu{\'e}, R~Schmidt, ST~Dawkins, and A~Rauschenbeutel.
\newblock Optical interface created by laser-cooled atoms trapped in the
  evanescent field surrounding an optical nanofiber.
\newblock {\em Phys. Rev. Lett.}, 104(20):203603, 2010.

\bibitem{goban12a}
A.~Goban, K.~S. Choi, D.~J. Alton, D.~Ding, C.~Lacro{\^u}te, M.~Pototschnig,
  T.~Thiele, N.~P. Stern, and H.~J. Kimble.
\newblock {Demonstration of a State-Insensitive, Compensated Nanofiber Trap}.
\newblock {\em Phys. Rev. Lett.}, 109:033603, 2012.

\bibitem{petersen14a}
Jan Petersen, J{\"u}rgen Volz, and Arno Rauschenbeutel.
\newblock Chiral nanophotonic waveguide interface based on spin-orbit
  interaction of light.
\newblock {\em Science}, 346(6205):67--71, 2014.

\bibitem{beguin14a}
J.-B. B\'eguin, E.~M. Bookjans, S.~L. Christensen, H.~L. S\o{}rensen, J.~H.
  M\"uller, E.~S. Polzik, and J.~Appel.
\newblock Generation and detection of a sub-poissonian atom number distribution
  in a one-dimensional optical lattice.
\newblock {\em Phys. Rev. Lett.}, 113:263603, Dec 2014.

\bibitem{sorensen16}
H.{\hspace{0.167em}}L. S{\o}rensen, J.-B. B{\'{e}}guin, K.{\hspace{0.167em}}W.
  Kluge, I.~Iakoupov, A.{\hspace{0.167em}}S. S{\o}rensen,
  J.{\hspace{0.167em}}H. Müller, E.{\hspace{0.167em}}S. Polzik, and J.~Appel.
\newblock Coherent backscattering of light off one-dimensional atomic strings.
\newblock {\em Physical Review Letters}, 117(13):133604, sep 2016.

\bibitem{corzo16}
Neil~V. Corzo, Baptiste Gouraud, Aveek Chandra, Akihisa Goban, Alexandra~S.
  Sheremet, Dmitriy~V. Kupriyanov, and Julien Laurat.
\newblock Large {Bragg} reflection from one-dimensional chains of trapped atoms
  near a nanoscale waveguide.
\newblock {\em Physical Review Letters}, 117(13):133603, sep 2016.

\bibitem{porras08a}
D~Porras and JI~Cirac.
\newblock {Collective generation of quantum states of light by entangled
  atoms}.
\newblock {\em Physical Review A}, 78(5):053816, 2008.

\bibitem{gonzaleztudela15a}
A.~Gonz\'alez-Tudela, V.~Paulisch, D.~E. Chang, H.~J. Kimble, and J.~I. Cirac.
\newblock Deterministic generation of arbitrary photonic states assisted by
  dissipation.
\newblock {\em Phys. Rev. Lett.}, 115:163603, Oct 2015.

\bibitem{gonzaleztudela16a}
A~Gonz{\'a}lez-Tudela, V~Paulisch, HJ~Kimble, and JI~Cirac.
\newblock Reliable multiphoton generation in waveguide qed.
\newblock {\em arXiv:1603.01243}, 2016.

\bibitem{chang12a}
DE~Chang, L~Jiang, AV~Gorshkov, and HJ~Kimble.
\newblock Cavity qed with atomic mirrors.
\newblock {\em New Journal of Physics}, 14(6):063003, 2012.

\bibitem{john90a}
Sajeev John and Jian Wang.
\newblock {Quantum electrodynamics near a photonic band gap: Photon bound
  states and dressed atoms}.
\newblock {\em Phys. Rev. Lett.}, 64:2418--2421, May 1990.

\bibitem{bykov75a}
Vladimir~P Bykov.
\newblock Spontaneous emission from a medium with a band spectrum.
\newblock {\em Soviet Journal of Quantum Electronics}, 4(7):861, 1975.

\bibitem{kurizki90a}
Gershon Kurizki.
\newblock Two-atom resonant radiative coupling in photonic band structures.
\newblock {\em Phys. Rev. A}, 42:2915--2924, Sep 1990.

\bibitem{douglas15a}
James~S Douglas, H~Habibian, C-L Hung, AV~Gorshkov, H~Jeff Kimble, and
  Darrick~E Chang.
\newblock Quantum many-body models with cold atoms coupled to photonic
  crystals.
\newblock {\em Nature Photonics}, 9(5):326--331, 2015.

\bibitem{gonzaleztudela15b}
Alejandro Gonz{\'a}lez-Tudela, C-L Hung, Darrick~E Chang, J~Ignacio Cirac, and
  HJ~Kimble.
\newblock Subwavelength vacuum lattices and atom--atom interactions in
  two-dimensional photonic crystals.
\newblock {\em Nature Photonics}, 9(5):320--325, 2015.

\bibitem{guimond16}
Pierre-Olivier Guimond, Alexandre Roulet, Huy~Nguyen Le, and Valerio Scarani.
\newblock Rabi oscillation in a quantum cavity: {Markov}ian and non-{Markov}ian
  dynamics.
\newblock {\em Physical Review A}, 93(2):023808, feb 2016.

\bibitem{shi15}
Tao Shi, Darrick~E. Chang, and J.~Ignacio Cirac.
\newblock Multiphoton-scattering theory and generalized master equations.
\newblock {\em Physical Review A}, 92(5):053834, nov 2015.

\bibitem{paulisch16}
V~Paulisch, H~J Kimble, and A~Gonz{\'{a}}lez-Tudela.
\newblock Universal quantum computation in waveguide {QED} using decoherence
  free subspaces.
\newblock {\em New Journal of Physics}, 18(4):043041, apr 2016.

\bibitem{SupMat}
See Appendix for i) a discussion on the atomic dynamics and ii) details of the
  protocol in the dissipative regime.

\bibitem{goban13a}
A.~Goban, C.-L. Hung, S.-P Yu, J.D. Hood, J.A. Muniz, J.H. Lee, M.J. Martin,
  A.C. McClung, K.S. Choi, D.E. Chang, O.~Painter, and H.J. Kimble.
\newblock Atom-light interactions in photonic crystals.
\newblock {\em Nat. Commun.}, 5:3808, 2014.

\bibitem{law96a}
C.~K. Law and J.~H. Eberly.
\newblock Arbitrary control of a quantum electromagnetic field.
\newblock {\em Phys. Rev. Lett.}, 76:1055--1058, Feb 1996.

\bibitem{sorensen03a}
Anders~S. S\o{}rensen and Klaus M\o{}lmer.
\newblock Measurement induced entanglement and quantum computation with atoms
  in optical cavities.
\newblock {\em Phys. Rev. Lett.}, 91:097905, Aug 2003.

\bibitem{albrecht16}
Andreas Albrecht, Tommaso Caneva, and Darrick~E. Chang.
\newblock Changing optical band structure with single photons.
\newblock {\em arXiv:1610.00988}, October 2016.

\bibitem{holstein40a}
T.~Holstein and H.~Primakoff.
\newblock {Field Dependence of the Intrinsic Domain Magnetization of a
  Ferromagnet}.
\newblock {\em Phys. Rev.}, 58:1098--1113, Dec 1940.

\bibitem{kurucz10}
Z.~Kurucz and K.~M\o{}lmer.
\newblock Multilevel holstein-primakoff approximation and its application to
  atomic spin squeezing and ensemble quantum memories.
\newblock {\em {Phys. Rev. A}}, 81:032314, Mar 2010.

\bibitem{dicke54a}
R.~H. Dicke.
\newblock {Coherence in Spontaneous Radiation Processes}.
\newblock {\em {Phys. Rev.}}, 93:99, 1954.

\bibitem{enk97}
S.~J. van Enk, J.~I. Cirac, and P.~Zoller.
\newblock Ideal quantum communication over noisy channels: A quantum optical
  implementation.
\newblock {\em Phys. Rev. Lett.}, 78(22):4293--4296, jun 1997.

\bibitem{borregaard15a}
J.~Borregaard, P.~K\'om\'ar, E.~M. Kessler, A.~S. S\o{}rensen, and M.~D. Lukin.
\newblock Heralded quantum gates with integrated error detection in optical
  cavities.
\newblock {\em Phys. Rev. Lett.}, 114:110502, Mar 2015.

\bibitem{fiurasek05a}
Jarom\'{\i}r Fiur\'a\ifmmode~\check{s}\else \v{s}\fi{}ek, Ra\'ul
  Garc\'{\i}a-Patr\'on, and Nicolas~J. Cerf.
\newblock Conditional generation of arbitrary single-mode quantum states of
  light by repeated photon subtractions.
\newblock {\em Phys. Rev. A}, 72:033822, Sep 2005.

\bibitem{motes16}
Keith~R. Motes, Ryan~L. Mann, Jonathan~P. Olson, Nicholas~M. Studer,
  E.~Annelise Bergeron, Alexei Gilchrist, Jonathan~P. Dowling, Dominic~W.
  Berry, and Peter~P. Rohde.
\newblock Efficient recycling strategies for preparing large fock states from
  single-photon sources --- applications to quantum metrology.
\newblock {\em Physical Review A}, 94(1):012344, March 2016.

\bibitem{gardiner_book00a}
G.~W. Gardiner and P.~Zoller.
\newblock {\em Quantum Noise: A Handbook of Markovian and Non-Markovian Quantum
  Stochastic Methods with Applications to Quantum Optics (Springer Series in
  Synergetics)}.
\newblock Springer-Verlag Berlin Heidelberg 2004, 2000.

\bibitem{dakna99a}
M.~Dakna, J.~Clausen, L.~Kn\"oll, and D.-G. Welsch.
\newblock Generation of arbitrary quantum states of traveling fields.
\newblock {\em Phys. Rev. A}, 59:1658--1661, Feb 1999.

\end{thebibliography}

\end{document}